\pdfoutput=1
\documentclass[prd,twocolumn,nofootinbib,superscriptaddress]{revtex4-2}

\usepackage{comment} 

\usepackage{graphicx}
\usepackage{amsmath,bm,amssymb,amsfonts,dsfont}
\usepackage[usenames,dvipsnames]{xcolor}
\usepackage[normalem]{ulem}
\usepackage{url}
\usepackage{array}
\usepackage{booktabs}
\usepackage{multirow}
\usepackage{float}
\usepackage[colorlinks  = true,
            linkcolor   = NavyBlue,
            urlcolor    = NavyBlue,
            citecolor   = NavyBlue,
            anchorcolor = NavyBlue]{hyperref}
\usepackage{cprotect}
\usepackage{mathtools,slashed}

\usepackage{bbold}

\usepackage{tikz-feynman} 
\tikzfeynmanset{every vertex/.style={/tikz/font=\small}}

\begin{document}

\title{Vector Higgs-Portal Dark Matter: How UV Completion Reopens Viable Parameter Space}

\author{Halim Shaikh}\email{halim.shaikh@tum.de}
\affiliation{Technical University of Munich, TUM School of Natural Sciences, Department of Physics,
James-Franck-Str. 1, 85748 Garching, Germany}

\author{Mattia Di Mauro}\email{dimauro.mattia@gmail.com}
\affiliation{Istituto Nazionale di Fisica Nucleare, Sezione di Torino, Via P. Giuria 1, 10125 Torino, Italy}

\date{\today}

\begin{abstract}
The particle nature of dark matter (DM) remains one of the central open problems in modern physics. Among the most extensively studied candidates are weakly interacting massive particles, whose parameter space is now under strong pressure from direct detection, indirect detection, and collider searches. In this work we revisit the Higgs-portal scenario with vector DM, first in an effective-field-theory description and then in a renormalizable UV-complete realization.
We show that the effective Higgs-portal model with a Proca vector coupled quadratically to the Standard Model Higgs is essentially excluded over almost all of its parameter space by current direct-detection limits, with only a narrow region near the Higgs resonance surviving with a required fine tuning of the DM to Higgs mass that should at the permille level. We then consider a UV completion based on an additional gauged \(U(1)_X\) symmetry, in which the DM candidate is a massive vector boson \(V\) and the scalar sector is extended by a dark Higgs that mixes with the Standard Model Higgs. In this framework, the presence of a second scalar mediator opens an additional resonant annihilation channel and can substantially weaken the direct-detection constraints. In particular, when the DM mass lies sufficiently close to the heavy-scalar resonance, \(m_V \simeq m_{H_2}/2\), the coupling required to reproduce the observed relic abundance can lie up to about two orders of magnitude below current direct-detection bounds, thereby opening viable parameter space that is absent in the effective description.
Our results highlight the importance of going beyond the effective-field-theory approximation in Higgs-portal vector DM models and show that UV-complete realizations can qualitatively change the phenomenological conclusions.
\end{abstract}

\maketitle

\section{Introduction}
\label{sec:intro}

A wide range of astrophysical and cosmological observations indicate that most of the matter content of the Universe consists of non-luminous and non-baryonic dark matter (DM)~\cite{Bertone:2016nfn,Bertone:2010zza,Cirelli:2024ssz}. 
A viable DM candidate must be cosmologically long-lived, electrically neutral, only weakly coupled to visible matter, non-relativistic by matter--radiation equality, and have sufficiently small self-interactions.
Within the Standard Model (SM) there is no particle that can simultaneously account for the inferred abundance and the required DM properties while satisfying laboratory constraints.
Therefore, if DM is a particle, its existence points to physics beyond the SM (BSM)~\cite{Bertone:2016nfn,Bertone:2010zza,Cirelli:2024ssz}.

Weakly Interacting Massive Particles (WIMPs) are a well-motivated class of candidates~\cite{Lee:1977ua,1978ApJ...223.1015G} that arise naturally in many BSM frameworks~\cite{WESS197439,1983PhRvL..50.1419G,Ellis:1983ew}.
In the standard thermal freeze-out picture, electroweak-scale masses and couplings generically lead to a relic density of the same order as the observed one, \(\Omega_{\rm DM} h^2 \simeq 0.12\)~\cite{Aghanim:2018eyx}.
This ``WIMP miracle''~\cite{Steigman:1984ac} has motivated extensive experimental programs in direct detection (DD)~\cite{Schumann:2019eaa}, collider searches~\cite{Boveia:2018yeb}, and indirect detection (ID)~\cite{Gaskins:2016cha}, complemented by precise cosmological measurements~\cite{Aghanim:2018eyx}.

Experimentally, the WIMP search program spans, at least, the following complementary directions.
DD experiments look for nuclear or electronic recoils induced by Galactic DM, reaching unprecedented sensitivities with multi-ton noble-liquid targets~\cite{Schumann:2019eaa,LZ:2024zvo,XENON:2023cxc}.
Collider searches aim to produce DM in high-energy SM particle collisions and infer its presence through missing-momentum signatures and invisible decays of SM states~\cite{Boveia:2018yeb,ABERCROMBIE2020100371,ATLAS:2020cjb,CMS:2018yfx,ATLAS:2022yvh,Baek:2012se}.
ID searches probe DM annihilation or decay products in rare cosmic messengers such as \(\gamma\) rays, antimatter, and neutrinos~\cite{Gaskins:2016cha,Fermi-LAT:2016afa}.
Interpreting these constraints consistently requires specifying a BSM model and, in particular, how the dark sector communicates with the SM.

The absence of conclusive signals so far has pushed many canonical WIMP realizations into increasingly constrained corners of parameter space, motivating both more systematic model building and more careful interpretations of null results~\cite{Arcadi:2017wqi,Cirelli:2024ssz}.
In particular, present DD experiments such as LZ and XENONnT exclude spin-independent (SI) WIMP--nucleon cross sections down to \(\mathcal{O}(10^{-48})\,\mathrm{cm}^2\) for weak-scale masses~\cite{LZ:2023,Aprile:2023XENONnT,LZ:2024zvo}.
In models where the same portal controls both freeze-out and scattering, these bounds can remove large portions of parameter space (see, e.g.,~\cite{Arcadi:2017kky,Arcadi:2019lka,DiMauro:2023tho,Arcadi:2024ukq,DiMauro:2025jia,Kong:2025ccv}).
In this regard, secluded DM models have been proposed as scenarios in which DM retains a thermal history while still evading stringent laboratory constraints~\cite{Pospelov:2007mp,DiMauro:2025jsb,DiMauro:2025uxt}.
Another possibility is provided by models in which the DD process arises only at loop level or is suppressed by a very small kinetic mixing (see, e.g.,~\cite{Koechler:2025ryv}).

From a theory standpoint, three complementary approaches are commonly used to describe DM interactions with the SM.
At energies well below the masses of new mediators, effective field theories (EFTs) provide a compact and economical parameterization of DM--SM interactions.
In this regime, the detailed mediator dynamics is integrated out and the phenomenology can often be summarized in terms of a small set of parameters, typically the DM mass and one or more effective couplings.
When the mediator can be produced on shell, or when it significantly affects the kinematics, simplified models that include the mediator explicitly become necessary~\cite{Abdallah:2015ter,ABERCROMBIE2020100371,Arina:2018zcq,DiMauro:2025jia,Arcadi:2024ukq,Koechler:2025ryv}.
Such models introduce at least the DM and mediator masses together with their couplings to the visible and dark sectors. 
Ultimately, in ultraviolet (UV)-complete theories---for instance supersymmetric extensions of the SM---gauge invariance and renormalizability fix the field content and interaction structure, and additional states can play an essential phenomenological role.
The main drawback is that UV completions often involve a larger number of particles and couplings, making a global exploration of their parameter space more challenging.

In this work we focus on Higgs-portal scenarios, where the SM Higgs doublet provides the dominant, and often minimal, connection between the visible and dark sectors.
Comprehensive discussions of Higgs-portal DM, including the interplay among relic density, invisible Higgs decays, and DD/ID searches, can be found, e.g., in Refs.~\cite{Silveira:1985rk,Arcadi:2019lka,Cline:2013gha,DiMauro:2023tho,Arcadi:2024ukq}.
A commonly used benchmark adds a singlet DM field and couples it to the Higgs bilinear.
For scalar DM, this yields a fully renormalizable setup, while for fermion and vector DM the simplest ``portal operators'' \cite{Beniwal:2015sdl,Arcadi:2019lka} are best viewed as low-energy descriptions of a more complete theory.
The singlet scalar Higgs-portal model---where the SM is extended by a single real scalar DM candidate \(S\)---is among the simplest renormalizable realizations: the operator \((H^\dagger H)S^2\) is the unique gauge-invariant, renormalizable interaction that couples \(S\) to the SM without introducing additional mediator fields.
See Ref.~\cite{DiMauro:2023tho} for a recent comprehensive study of its phenomenology and of the interplay among relic density, collider bounds, DD, and ID constraints where the authors have demonstrated that only in the region where the DM mass is about half of the Higgs one can survive.

The vector case is particularly instructive and has been studied extensively~\cite{Ko:2014gha,Beniwal:2015sdl,Arcadi:2020jqf,Arcadi:2024ukq}.
An effective interaction of the form \((H^\dagger H)\,V_\mu V^\mu\) is widely employed to interpret limits from invisible Higgs decays and to map them onto SI scattering cross sections.
However, a massive spin-1 field described purely as a Proca state does not constitute a consistent UV completion: perturbative unitarity and gauge invariance typically require additional dynamics, e.g.\ a dark gauge symmetry broken by a dark Higgs field, and the resulting extra degrees of freedom can spoil the naive one-to-one correspondence between collider and DD constraints~\cite{Kahlhoefer:2015bea,Englert:2016joy}.
In UV-complete realizations, mixing in the scalar sector can induce interference patterns in both annihilation and scattering amplitudes, potentially opening regions of parameter space that remain compatible with current laboratory constraints~\cite{Duch:2015jta,Arcadi:2020jqf}.

Motivated by these considerations, we study a renormalizable Higgs-portal framework in which the DM is the stable gauge boson of a dark Abelian symmetry that is spontaneously broken by an additional scalar field.
After symmetry breaking, the portal to the SM is mediated by two CP-even scalar mass eigenstates, whose mixing controls Higgs-signal-strength modifications, invisible decays, and the coupling of DM to nuclei.
We perform a combined and up-to-date analysis of the parameter space, incorporating (i) the relic abundance computed from the full Boltzmann evolution in the thermal freeze-out regime~\cite{Kolb:1990vq}, (ii) current and future DD bounds~\cite{LZ:2024zvo,XENON:2023cxc,DARWIN:2016hyl}, (iii) LHC constraints from invisible Higgs decays~\cite{ATLAS:2020cjb,CMS:2018yfx,ATLAS:2022yvh}, and (iv) bounds from ID using the combined {\it Fermi}-LAT observations of Milky Way dwarf spheroidal galaxies (dSphs)~\cite{McDaniel:2023bju}.
We also discuss how the EFT limit emerges, or fails to emerge, in controlled decoupling regimes, thereby clarifying when collider and astroparticle constraints can be compared in a meaningful way~\cite{Arcadi:2020jqf,Englert:2016joy}.

The paper is organized as follows.
In Sec.~\ref{sec:eft} we present the EFT model Lagrangian and its main interactions, discuss its regime of validity, and derive the constraints from the relic density together with collider, DD, and ID searches.
In Sec.~\ref{sec:UV} we present the corresponding analysis for the UV-complete model.
Finally, Sec.~\ref{sec:conclusions} contains our conclusions.

\section{The EFT Higgs-portal vector dark matter model}
\label{sec:eft}

\subsection{Lagrangian, particle content and interactions}

We start from the EFT description of a neutral spin-1 DM candidate \(V_\mu\) coupled to the SM exclusively through the Higgs portal.
At the EFT level, \(V_\mu\) is treated as a massive Proca field, and its stability is ensured by imposing a \(\mathbb{Z}_2\) symmetry \(V_\mu \to -V_\mu\), which forbids linear and cubic operators.
This setup is widely used as a benchmark for collider, DD, and ID searches, and provides a convenient limiting description of simple UV completions based on a broken dark gauge symmetry \cite{Arcadi:2020jqf,Beniwal:2015sdl,Duch:2015jta,Ko:2014gha}.

The EFT Lagrangian can be written as
\begin{widetext}
\begin{equation}
\mathcal{L}
=
\mathcal{L}_{\rm SM}
-\frac{1}{4}\,V_{\mu\nu}V^{\mu\nu}
+\frac{1}{2}\,\mu_V^{2}\,V_\mu V^\mu
-\frac{\lambda_V}{4}\left(V_\mu V^\mu\right)^2
+\frac{\lambda_{HV}}{2}\,V_\mu V^\mu\,H^\dagger H \,,
\label{eq:eft_lagrangian}
\end{equation}
\end{widetext}
where \(V_{\mu\nu}\equiv \partial_\mu V_\nu-\partial_\nu V_\mu\) is the field-strength tensor, \(H\) is the SM Higgs doublet, \(\lambda_{HV}\) is the Higgs--vector portal coupling, and \(\lambda_V\) parametrizes a quartic self-interaction.
This last term does not play a central role in the relic-density and detection phenomenology and will be neglected in the following when appropriate \cite{Arcadi:2020jqf}.

After electroweak symmetry breaking, in unitary gauge,
\begin{equation}
H=\frac{1}{\sqrt{2}}
\begin{pmatrix}
0\\ v+h
\end{pmatrix},
\qquad
H^\dagger H=\frac{1}{2}\left(v^2+2vh+h^2\right),
\label{eq:Hexpansion}
\end{equation}
with \(v\simeq 246~{\rm GeV}\).
The portal term induces both a contribution to the physical vector mass and the interactions relevant for annihilation, DD, and invisible Higgs decays:
\begin{align}
\mathcal{L}
&\supset
\frac{1}{2}m_V^2\,V_\mu V^\mu
+\frac{\lambda_{HV}v}{2}\,h\,V_\mu V^\mu
+\frac{\lambda_{HV}}{4}\,h^2\,V_\mu V^\mu,
\label{eq:eft_after_ewsb}
\end{align}
where the DM mass is given by
\begin{equation}
m_V^2 = \mu_V^2+\frac{\lambda_{HV}}{2}\,v^2.
\end{equation}
Thus, for phenomenology it is convenient to trade \(\mu_V\) for the physical mass \(m_V\), so that, up to the usually irrelevant terms proportional only to \(\lambda_V\), the EFT is effectively characterized by the two parameters \(\{m_V,\lambda_{HV}\}\) \cite{Arcadi:2020jqf,Beniwal:2015sdl}.

At momentum transfers relevant for DD, \( |q^2| \ll m_h^2 \), where \(m_h\) is the Higgs mass, the Higgs propagator can be contracted to a point and the Higgs field can be integrated out.
Starting from the Higgs--portal interaction \((\lambda_{HV}/2)\,V_\mu V^\mu\,H^\dagger H\), electroweak symmetry breaking generates the trilinear coupling
\begin{equation}
\mathcal{L}\supset \frac{\lambda_{HV} v}{2}\,h\,V_\mu V^\mu,
\end{equation}
which, combined with the Higgs couplings to quark scalar currents and to gluons through the QCD trace anomaly, induces the following low-energy effective interaction between the vector DM field and light quarks and gluons:
\begin{equation}
\mathcal{L}^{\rm eff}_{V}
=
-\frac{\lambda_{HV}}{4m_h^2}\,V_\mu V^\mu
\left[
\sum_{q=u,d,s} m_q\,\bar q q
-\frac{\alpha_s}{4\pi}\,G^a_{\mu\nu}G^{a\,\mu\nu}
\right].
\label{eq:Leff_vector}
\end{equation}
Here \(m_q\) are the SM light-quark current masses, \(\alpha_s\equiv g_s^2/(4\pi)\), and \(G^a_{\mu\nu}\) is the gluon field-strength tensor.
Eq.~\eqref{eq:Leff_vector} is the starting point for deriving the SI DM--nucleon coupling after taking nucleon matrix elements of the scalar quark and gluon operators.

\subsection{Range of validity of the EFT approach}
\label{sec:validity}

The EFT in Eq.~\eqref{eq:eft_lagrangian} treats \(V_\mu\) as a massive Proca field.
In this description, the longitudinal polarization behaves at energies \(E\gg m_V\) as \(\epsilon_L^\mu(p)\sim p^\mu/m_V\).
As a consequence, scattering amplitudes involving external longitudinal vectors can grow with energy.
In particular, the Higgs-portal interaction induces contributions to \(VV\to VV\) (and related channels) whose amplitude scales parametrically as
\begin{equation}
\mathcal{M}(V_LV_L\to V_LV_L)\ \sim\ \lambda_{HV}\,\frac{E^2}{m_V^2}\,,
\end{equation}
up to angular factors.
This energy growth signals that the EFT cannot be extrapolated indefinitely: above a certain cutoff the theory violates perturbative unitarity and must be completed by additional degrees of freedom, e.g.\ a dark gauge symmetry with a dark Higgs, that restore good high-energy behavior.

A conservative way to delineate the domain of validity is to require perturbative unitarity of the \(J=0\) partial wave, \(|a_0|\lesssim 1/2\).
Using \(a_0\sim \mathcal{M}/(16\pi)\), one finds that the EFT must satisfy
\begin{equation}
\lambda_{HV}\,\frac{E^2}{m_V^2}\ \lesssim\ 16\pi
\,\,\Rightarrow\,\,
E \ \lesssim\ \Lambda_{\rm uni}\ \equiv\ \frac{\sqrt{16\pi}}{\sqrt{\lambda_{HV}}}\,m_V\,.
\label{eq:unitarity_cutoff}
\end{equation}
Thus, for fixed \(\lambda_{HV}\), an overly light vector implies a low unitarity cutoff, making the EFT unreliable already near the electroweak scale.

In Higgs-portal phenomenology it is common to adopt an additional, model-independent consistency criterion by demanding that the cutoff not lie below the electroweak scale, i.e.\ \(\Lambda_{\rm uni}\gtrsim v\).
This yields the parametric bound
\begin{equation}
m_V \gtrsim \sqrt{\frac{\lambda_{HV}}{16\pi}}\,v\,,
\label{eq:unitarity_bound}
\end{equation}
which excludes the regime of simultaneously small \(m_V\) and large \(\lambda_{HV}\).
This should be viewed as an indicator of where the Proca EFT ceases to be self-consistent; in a UV-complete gauge theory the additional states (dark Higgs, etc.) cure the bad high-energy behavior and can modify the correlations among relic density, DD, and collider constraints \cite{Arcadi:2020jqf}.

\medskip

\subsection{Annihilation cross sections into SM final states}
\label{subsec:annihilation_allSM}


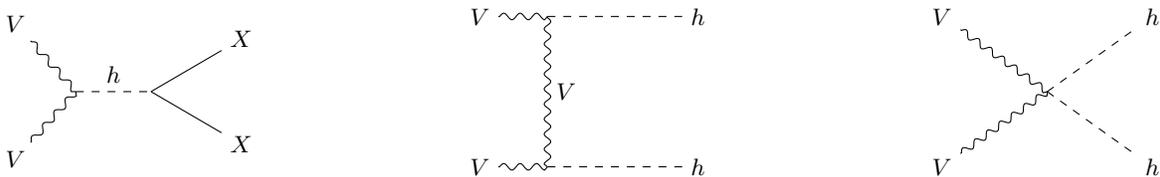
\begin{figure*}[t]
\centering

\begin{minipage}{0.32\textwidth}
\centering
\begin{tikzpicture}
\begin{feynman}
  \vertex (V1) at (-1.6,  0.9) {$V$};
  \vertex (V2) at (-1.6, -0.9) {$V$};
  \vertex (a)  at (-0.8,  0.0);
  \vertex (b)  at ( 0.2,  0.0);
  \vertex (x1) at ( 1.4,  0.7) {$X$};
  \vertex (x2) at ( 1.4, -0.7) {$X$};

  \diagram*{
    (V1) -- [boson] (a),
    (V2) -- [boson] (a),
    (a)  -- [scalar, dashed, edge label={$h$}] (b),
    (b)  -- [plain] (x1),
    (b)  -- [plain] (x2),
  };
\end{feynman}
\end{tikzpicture}
\end{minipage}
\hfill
\begin{minipage}{0.32\textwidth}
\centering
\begin{tikzpicture}
\begin{feynman}
  \vertex (V1) at (-2.0,  1.0) {$V$};
  \vertex (V2) at (-2.0, -1.0) {$V$};
  \vertex (a)  at (-1.1,  1.0);
  \vertex (b)  at (-1.1, -1.0);
  \vertex (h1) at ( 0.9,  1.0) {$h$};
  \vertex (h2) at ( 0.9, -1.0) {$h$};

  \diagram*{
    (V1) -- [boson] (a) -- [scalar, dashed] (h1),
    (V2) -- [boson] (b) -- [scalar, dashed] (h2),
    (a)  -- [boson, edge label={$V$}] (b),
  };
\end{feynman}
\end{tikzpicture}
\end{minipage}
\hfill
\begin{minipage}{0.32\textwidth}
\centering
\begin{tikzpicture}
\begin{feynman}
  \vertex (V1) at (-1.4,  1.0) {$V$};
  \vertex (V2) at (-1.4, -1.0) {$V$};
  \vertex (h1) at ( 1.4,  1.0) {$h$};
  \vertex (h2) at ( 1.4, -1.0) {$h$};
  \vertex (x)  at ( 0.0,  0.0);

  \diagram*{
    (V1) -- [boson] (x) -- [scalar, dashed] (h1),
    (V2) -- [boson] (x) -- [scalar, dashed] (h2),
  };
\end{feynman}
\end{tikzpicture}
\end{minipage}

\caption{Tree-level annihilation topologies for vector DM \(V\) in the Higgs-portal setup.
Left: \(s\)-channel Higgs exchange, \(VV\to h^\ast\to XX\), where \(X\) denotes any kinematically accessible SM final state
(e.g.\ \(f\bar f\), \(W^+W^-\), \(ZZ\), or \(hh\)). Middle: \(t/u\)-channel \(V\) exchange contributing to \(VV\to hh\).
Right: contact interaction \(VVhh\) arising after electroweak symmetry breaking.}
\label{fig:VHP_tree_substituted}
\end{figure*}

\begin{figure*}[t]
\centering
\begin{tikzpicture}
\begin{feynman}
  \vertex (V1) at (-2.4,  1.1) {$V$};
  \vertex (V2) at (-2.4, -1.1) {$V$};
  \vertex (a)  at (-1.5,  0.0);

  \vertex (h)  at (-0.6,  0.0);

  \vertex (tL) at ( 0.2,  0.0);   
  \vertex (tU) at ( 1.2,  0.9);   
  \vertex (tD) at ( 1.2, -0.9);   

  \vertex (g1) at ( 2.3,  0.9) {$g$};
  \vertex (g2) at ( 2.3, -0.9) {$g$};

  \diagram*{
    (V1) -- [boson] (a),
    (V2) -- [boson] (a),
    (a)  -- [scalar, dashed, edge label={$h$}] (h),

    (h)  -- [plain] (tL),

    (tL) -- [fermion, edge label={$q$}] (tU),
    (tU) -- [fermion, edge label={$q$}] (tD),
    (tD) -- [fermion, edge label={$q$}] (tL),

    (tU) -- [gluon] (g1),
    (tD) -- [gluon] (g2),
  };
\end{feynman}
\end{tikzpicture}

\caption{Loop-induced annihilation \(VV\to gg\) mediated by an off-shell Higgs \(h^\ast\) through a quark loop.}
\label{fig:VV_to_gg_loop}
\end{figure*}
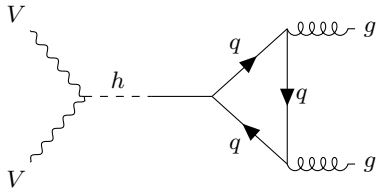

The relic density and ID signals are controlled by \(VV\) annihilation into SM final states, dominantly through an \(s\)-channel Higgs exchange.
It is convenient to express the channels generated purely by \(s\)-channel Higgs exchange in terms of the off-shell Higgs partial widths evaluated at virtuality \(\sqrt{s}\), \(\Gamma_{h^\ast\to X}(s)\), which are well determined theoretically and tested experimentally~\cite{Arcadi:2020jqf,Beniwal:2015sdl,LHCHXSWG:2013,ATLAS-CMS:2016couplings}.
For an initial state of two identical massive vectors of mass \(m_V\), the tree-level annihilation cross section into an arbitrary final state \(X\) mediated only by a scalar propagator can be written as
\begin{widetext}
\begin{equation}
\sigma(VV\to X)(s)=
\frac{\lambda_{HV}^2\,v^2}{m_V^4}\;
\frac{s^2-4sm_V^2+12m_V^4}{72\pi\,s}\;
\frac{\sqrt{s}}{\sqrt{s-4m_V^2}}\;
\frac{\Gamma_{h^\ast\to X}(s)}
{\left(s-m_h^2\right)^2+m_h^2\left[\Gamma_h^{\rm tot}(s)\right]^2}\,,
\label{eq:sigmavv_general_width}
\end{equation}
\end{widetext}
where \(\Gamma_{h^\ast\to X}(s)\) is the partial width of a virtual Higgs boson \(h^\ast\) of invariant mass \(\sqrt{s}\) into the final state \(X\), and \(\Gamma_h^{\rm tot}(s)=\Gamma_{h,\rm SM}(s)+\Gamma_{VV}(s)\) is the total Higgs width including the invisible channel when open (cf.\ Sec.~\ref{subsubsec:higgs_invisible_width}).
Eq.~\eqref{eq:sigmavv_general_width} is particularly convenient because it automatically covers fermions, electroweak gauge bosons, and loop-induced channels such as \(gg\)~\cite{Arcadi:2020jqf,Beniwal:2015sdl}.
We show in Figs.~\ref{fig:VHP_tree_substituted} and~\ref{fig:VV_to_gg_loop} the Feynman diagrams for the dominant annihilation channels relevant for the model.

\paragraph{Fermion pairs.}
For \(X=f\bar f\) (color factor \(N_c\)), one may equivalently use
\begin{widetext}
\begin{equation}
\sigma(VV\to f\bar f)
=
\frac{\lambda_{HV}^2\,m_f^2\,N_c}{288\pi\,m_V^4\,s}\,
\frac{(s-4m_f^2)^{3/2}}{(s-4m_V^2)^{1/2}}\,
\frac{s^2-4sm_V^2+12m_V^4}
{(s-m_h^2)^2+m_h^2\left[\Gamma_h^{\rm tot}(s)\right]^2}\,,
\label{eq:sigmavv_ff}
\end{equation}
\end{widetext}
which follows from Eq.~\eqref{eq:sigmavv_general_width} upon inserting
\(\Gamma_{h^\ast\to f\bar f}(s)\).

\paragraph{Electroweak gauge bosons and Higgs pairs.}
For \(X=WW\) and \(ZZ\), the same master formula applies with the corresponding off-shell partial widths \(\Gamma_{h^\ast\to WW}(s)\) and \(\Gamma_{h^\ast\to ZZ}(s)\)~\cite{Djouadi:2005gj,Arcadi:2020jqf}.
These channels dominate for \(m_V\) above the electroweak scale and exhibit characteristic threshold features when \(m_V\) approaches \(m_W\) and \(m_Z\).
In particular, it is important to include off-shell contributions near threshold.
The decay widths of the electroweak gauge bosons are \(\Gamma_W\simeq 2.1~\mathrm{GeV}\) and \(\Gamma_Z\simeq 2.5~\mathrm{GeV}\), so that near threshold the \(VV\to WW^\ast\) and \(VV\to ZZ^\ast\) processes can be non-negligible.
To include these effects, we evaluate the annihilation into four-fermion final states, \(VV\to WW^{(\ast)}\to 4f\) and \(VV\to ZZ^{(\ast)}\to 4f\), following the procedure described in Ref.~\cite{DiMauro:2023tho}.

For \(X=hh\), the full tree-level amplitude receives contributions not only from the \(s\)-channel Higgs exchange, but also from the contact interaction \(VVhh\) and from the \(t/u\)-channel exchange of the vector DM particle, as shown in Fig.~\ref{fig:VHP_tree_substituted}.
Therefore, the \(hh\) channel is treated separately in the numerical implementation with the full set of tree-level diagrams.

\paragraph{Gluons (loop-induced).}
For \(X=gg\), the annihilation proceeds through the loop-induced effective
\(hgg\) coupling (dominated by the top quark in the SM) and is consistently
included by using \(\Gamma_{h^\ast\to gg}(s)\) in Eq.~\eqref{eq:sigmavv_general_width}
\cite{Djouadi:2005gj,Arcadi:2020jqf}.

\begin{figure}[t]
\centering
\includegraphics[width=0.99\linewidth]{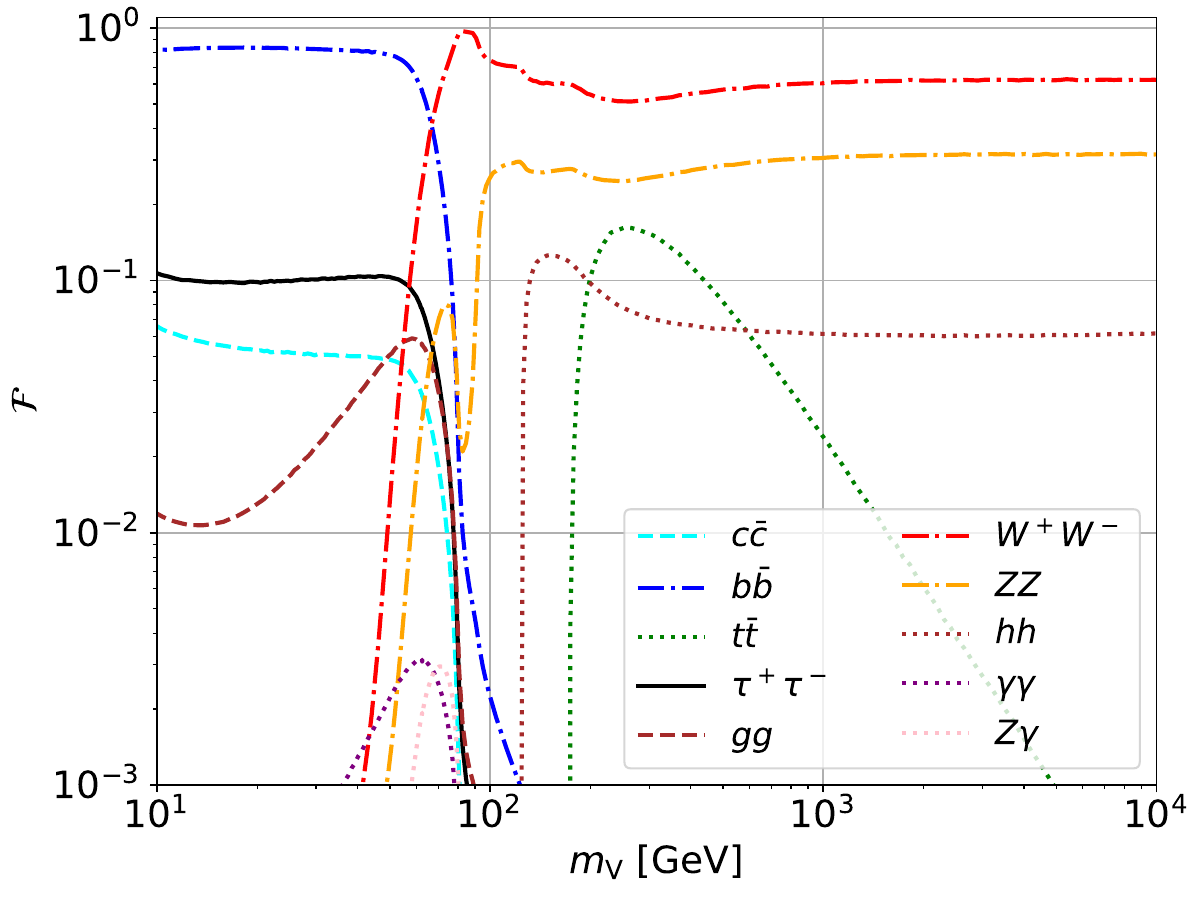}
\caption{Fractional contribution \(\mathcal{F}_i=\langle\sigma v\rangle_i/\langle\sigma v\rangle_{\rm tot}\)
of each annihilation channel as a function of the DM mass, computed for a fixed \(\lambda_{HV}=0.01\).}
\label{fig:Br}
\end{figure}

The fractional contribution of each channel,
\begin{equation}
\mathcal{F}_i(m_V)\equiv
\frac{\langle\sigma v\rangle_i}{\langle\sigma v\rangle_{\rm tot}}\,,
\end{equation}
is shown in Fig.~\ref{fig:Br} as a function of the DM mass for \(\lambda_{HV}=0.01\).
For \(m_V\lesssim m_W\) the dominant contribution typically comes from the heaviest
kinematically accessible fermions (notably \(b\bar b\)), while above the electroweak thresholds the \(W^+W^-\) and \(ZZ\) channels rapidly take over, with \(hh\) becoming important once \(m_V\gtrsim m_h\).
Off-shell effects (e.g.\ \(WW^\ast\)) can make the gauge-boson channels relevant even slightly below the on-shell threshold. In fact, we see in Fig.~\ref{fig:Br} that the annihilation channel into $WW^*$ and $ZZ^*$ can provide the largest cross section even a few GeV below the propagator masses.

In the numerical implementation we include all kinematically open channels.
The \(f\bar f\), \(WW\), \(ZZ\), and \(gg\) final states are evaluated through Eq.~\eqref{eq:sigmavv_general_width}, while the \(hh\) final state is computed from the full tree-level amplitude including the \(s\)-channel, contact, and \(t/u\)-channel contributions.
Eq.~\eqref{eq:sigmavv_ff} is recovered as a cross-check in the fermionic case.
All rates entering the thermal average \(\langle\sigma v\rangle\) are evaluated at tree level, with \(gg\) included via the leading SM loop-induced width.
For the indirect-detection analysis we use \texttt{micrOMEGAs} and \texttt{MadDM}~\cite{Belanger:2013oya,Beniwal:2015sdl} to compute the annihilation rates and branching fractions, and we compare the resulting predictions with the bounds derived from {\it Fermi}-LAT observations of dSphs reported in Ref.~\cite{McDaniel:2023bju}.
Our estimate of the $\gamma$-ray flux is based on the {\tt CosmiXs} source spectra~\cite{Arina:2023eic,DiMauro:2024kml}.

\subsection{Relic density}
\label{sec:RD}

The present-day DM abundance is precisely measured by the \emph{Planck} satellite, $\Omega_{\rm DM} h^2 = 0.120$ with $\mathcal{O}(1\%)$ uncertainty \cite{Planck:2018vyg}.
Computing the relic abundance from first principles amounts to following the phase-space distribution of the DM particle in an expanding Friedmann--Robertson--Walker Universe. For vector DM $V$, this is described by the Boltzmann equation for the phase-space density $f_V(\vec p,t)$
\cite{Kolb:1990vq,Edsjo:1997bg}:
\begin{equation}
E\left(\partial_t - H\,\vec p\cdot\nabla_{\vec p}\right)f_V(\vec p,t)
= \mathcal{C}\!\left[f_V\right],
\label{eq:RD1}
\end{equation}
where $E$ and $\vec p$ are the energy and momentum of $V$, $H$ is the Hubble rate, and $\mathcal{C}$ is the collision operator encoding interactions with the SM bath. The collision term can be decomposed schematically into elastic and inelastic contributions, $\mathcal{C}=\mathcal{C}_{\rm el}+\mathcal{C}_{\rm ann}$,
where elastic scatterings maintain kinetic equilibrium, while number-changing processes (dominated here by annihilation) control chemical equilibrium.
A detailed discussion of the structure and impact of $\mathcal{C}_{\rm el}$ and $\mathcal{C}_{\rm ann}$ can be found in Ref.~\cite{Binder:2017rgn}.

\medskip

For the standard thermal freeze-out of weakly interacting massive particles, Eq.~\eqref{eq:RD1} is commonly simplified by assuming that DM remains in kinetic equilibrium with the SM plasma throughout chemical decoupling. Under this assumption the DM distribution stays close to a thermal form,
$f_V(\vec p,t)\propto f_{V,{\rm eq}}(\vec p,t)$, and the evolution reduces to an ordinary differential equation for the number density. One obtains the well-known Zeldovich--Okun--Pikelner / Lee--Weinberg equation
\cite{YaBZel'dovich_1966,PhysRevLett.39.165}:
\begin{equation}
\frac{d n_V}{dt}+3Hn_V
=
-\langle\sigma v_{\rm Mol}\rangle_T
\left(n_V^2-n_{V,{\rm eq}}^2\right),
\label{eq:RD2}
\end{equation}
with $n_V=g_V\int d^3p/(2\pi)^3\,f_V(p)$ and $g_V=3$ for a massive spin-1 particle.
In the non-relativistic regime, $f_{V,{\rm eq}}$ is well approximated by a Maxwell--Boltzmann distribution. The thermal average can be expressed in terms of the center-of-mass energy $s$ as
\begin{equation}
\langle\sigma v\rangle_T
=
\int_{4m_V^2}^{\infty} ds\;
\frac{s\,\sqrt{s-4m_V^2}\,K_1(\sqrt{s}/T)}
{16\,T\,m_V^4\,K_2^2(m_V/T)}\,
(\sigma v)(s),
\label{eq:thermal_average}
\end{equation}
where $K_i$ are modified Bessel functions, $\sigma$ describes the DM annihilation cross section in Eq.~\ref{eq:sigmavv_general_width} and $v$ is the Moller DM relative velocity.

\medskip

The assumption of kinetic equilibrium at the time of the chemical decoupling is often justified because elastic scattering off abundant light SM particles can be much faster than the annihilation rate around freeze-out. However, this argument can fail when annihilation proceeds through a narrow $s$-channel resonance such as the Higgs:
(i) annihilation can be resonantly enhanced while elastic scattering is not, and 
(ii) Higgs couplings to light SM states are Yukawa suppressed, reducing the efficiency of elastic momentum exchange with the thermal bath. As a result, kinetic decoupling may occur during chemical freeze-out, invalidating the
reduction to Eq.~\eqref{eq:RD2} and requiring the full phase-space treatment including $\mathcal{C}_{\rm el}$ \cite{Binder:2017rgn}.

In practice, for the Higgs-portal vector DM scenario considered here, the impact of this effect is modest. Ref.~\cite{DiMauro:2025jia} showed that including kinetic decoupling modifies the relic abundance by at most
$\mathcal{O}(20\%)$, and only in a narrow region where $m_V$ lies within a few percent of the Higgs resonance, $m_V\simeq m_h/2$. Given that this correction is both localized and subdominant for the parameter space of interest, we adopt
the standard treatment based on Eq.~\eqref{eq:RD2} in the remainder of this work, and use the full kinetic-decoupling computation only as a cross-check in the resonant regime.
See Ref.~\cite{Giorgio26} for an exact calculation of the relic density obtained by solving the full Boltzmann equation.

For the numerical evaluation of the DM relic density we emply the \texttt{micrOMEGAs} numerical code \cite{Belanger:2013oya}.

\subsection{Invisible Higgs decay width}
\label{subsubsec:higgs_invisible_width}

If $m_V<m_h/2$, the SM-like Higgs boson can decay invisibly into a pair of
dark vectors, $h\to VV$. For an on-shell Higgs with mass $m_h$, the partial
width is
\begin{equation}
\Gamma(h\to VV)
=
\frac{\lambda_{HV}^2\,v^2\,m_h^3}{128\pi\,m_V^4}\,
\beta_{Vh},
\label{eq:h_to_VV_width}
\end{equation}
with
\begin{equation}
\beta_{Vh}\equiv
\left(1-\frac{4m_V^2}{m_h^2}+\frac{12m_V^4}{m_h^4}\right)
\sqrt{1-\frac{4m_V^2}{m_h^2}}\,,
\label{eq:h_to_VV_beta}
\end{equation}
in agreement with standard Higgs-portal results
\cite{Arcadi:2020jqf,Beniwal:2015sdl}.

When needed---for instance when the Higgs propagator is probed close to
resonance and/or close to the kinematic threshold $s\simeq 4m_V^2$---it is
convenient to use an $s$-dependent (off-shell) partial width,
\begin{equation}
\Gamma_{VV}(s)
=
\frac{\lambda_{HV}^2\,v^2}{128\pi\,s\,m_V^4}\,
\sqrt{s-4m_V^2}\,
\left(12m_V^4+s^2-4s\,m_V^2\right),
\label{eq:hstar_to_VV_width}
\end{equation}
which reduces to Eq.~\eqref{eq:h_to_VV_width} for $s=m_h^2$.
In an improved treatment of resonant amplitudes one may then employ an
$s$-dependent Higgs total width in the propagator,
\begin{equation}
\Gamma_{h,\rm tot}(s)=\Gamma_{h,\rm SM}(s)+\Gamma_{VV}(s)\,,
\end{equation}
whenever the invisible channel is kinematically open.

\medskip

Collider constraints from invisible Higgs decays provide a robust probe of this
scenario, being largely independent of astrophysical systematics (e.g.~the DM Galactic density).
For $m_V<m_h/2$, the new channel contributes to the Higgs invisible width,
$\Gamma_{h,\rm inv}\equiv \Gamma(h\to VV)$, and the invisible branching ratio is
\begin{equation}
\mathcal{B}_{h,\rm inv}=
\frac{\Gamma_{h,\rm inv}}
{\Gamma_{h,\rm SM}+\Gamma_{h,\rm inv}}\,.
\label{eq:BR_hinv}
\end{equation}
The latest ATLAS constraint gives a $95\%$~CL upper limit
$\mathcal{B}_{h,\rm inv}<0.145$ \cite{ATLAS:2022yvh}.

In this work we are particularly interested in the resonant region
$m_V\simeq m_h/2$, where threshold effects and the rapid turn-on of
$\Gamma(h\to VV)$ can make a fixed-width approximation less accurate.
As emphasized in Ref.~\cite{Heisig:2019vcj}, a consistent reinterpretation may
require using an $s$-dependent total width in the Higgs propagator.
One can implement this following, e.g., the procedure of
Refs.~\cite{Heisig:2019vcj,DiMauro:2023tho,Baek:2014jga}.
For the parameter ranges considered here, we verified that this refinement
does not materially change the inferred exclusion compared to the simpler
on-shell treatment. We therefore impose the invisible-decay bound by comparing
the theoretical prediction of Eq.~\eqref{eq:BR_hinv} (with
$\Gamma_{h,\rm inv}=\Gamma(h\to VV)$) to the ATLAS limit \cite{ATLAS:2022yvh}.

\subsection{Direct Detection searches}
\label{sec:DD}

Underground DD experiments search for nuclear recoils induced by the elastic scattering of DM particles off target nuclei. The most stringent bounds on the SI DM--nucleon cross section currently come from dual-phase xenon time-projection chambers such as XENONnT and LZ \cite{XENON:2023sxq,LZ:2022ufs}. Experimental limits are typically
reported as $90\%$~CL upper bounds on $\sigma_{\rm SI}$ as a function of the DM mass under the Standard Halo Model (SHM) assumptions for the local phase-space distribution. In this framework one adopts a local DM density
$\rho_\odot\simeq 0.3~{\rm GeV\,cm^{-3}}$ and a Maxwellian velocity distribution
with characteristic speed $v_0\simeq 220~{\rm km\,s^{-1}}$, escape speed
$v_{\rm esc}\simeq 544~{\rm km\,s^{-1}}$, and Earth speed
$v_E\simeq 232~{\rm km\,s^{-1}}$ (see, e.g., the experimental analyses in
Refs.~\cite{XENON:2023sxq,LZ:2022ufs}). With these assumptions, current limits
reach $\sigma_{\rm SI}\sim \mathcal{O}(10^{-48})\,{\rm cm}^2$ around DM masses of a few
tens of GeV, and future improvements, e.g.~with the DARWIN experiment, are expected to push the sensitivity
toward the $\sim \mathcal{O}(10^{-49})\,{\rm cm}^2$ level \cite{DARWIN:2016hyl}, i.e.~within the reach of the neutrino floor cross sections.

\medskip

In the Higgs-portal vector DM model considered here, scattering off nuclei is dominated by $t$-channel Higgs exchange and is therefore purely SI and approximately isospin conserving. The relevant interaction after electroweak symmetry breaking can be written as
\begin{equation}
\mathcal{L}\supset \frac{1}{2}\,g_{hVV}\,h\,V_\mu V^\mu,
\qquad
g_{hVV}=\lambda_{HV}\,v,
\label{eq:ghVV_def}
\end{equation}
where the last equality follows from our convention for the Higgs-portal
coupling (consistent with the decay width and annihilation expressions used in
Secs.~\ref{subsubsec:higgs_invisible_width} and \ref{subsec:annihilation_allSM}).
The Higgs coupling to nucleons is parameterized as
\begin{equation}
\mathcal{L}\supset g_{hNN}\,h\,\bar N N,
\qquad
g_{hNN}=\frac{f_N\,m_N}{v},
\label{eq:ghNN_def}
\end{equation}
where $m_N$ is the nucleon mass and $f_N$ encodes the scalar matrix elements
of the nucleon. A commonly used reference value is $f_N\simeq 0.30$
\cite{Cline:2013gha}.

In the zero-momentum-transfer limit, the resulting SI DM--nucleon cross section
is
\begin{equation}
\sigma_{\rm SI}^{(N)}
=
\frac{g_{hVV}^2\,g_{hNN}^2}{4\pi\,m_h^4}\,
\frac{m_N^2}{(m_V+m_N)^2}
=
\frac{\lambda_{HV}^2\,f_N^2}{4\pi\,m_h^4}\,
\frac{m_N^4}{(m_V+m_N)^2},
\label{eq:sigmaSI_vector}
\end{equation}
where $m_V$ is the vector DM mass and $m_h$ the Higgs mass. It is sometimes
useful to rewrite Eq.~\eqref{eq:sigmaSI_vector} in terms of the reduced mass
$\mu_{VN}=m_V m_N/(m_V+m_N)$,
\begin{equation}
\label{eq:sigma_eft_final}
\sigma_{\rm SI}^{(N)}
=
\frac{\lambda_{HV}^2\,f_N^2}{4\pi\,m_h^4}\,
\frac{\mu_{VN}^2\,m_N^2}{m_V^2}.
\end{equation}

\medskip

For the numerical evaluation of $\sigma_{\rm SI}$ we employ both
\texttt{MadDM} \cite{Backovic:2015cra} and \texttt{micrOMEGAs}
\cite{Belanger:2013oya}. We find agreement at the few-percent level across the relevant mass range. Residual differences can be traced to the treatment of QCD running and to additional subleading contributions (e.g.\ twist-2 operators) included by default in \texttt{micrOMEGAs} but not in a minimal tree-level implementation.

\subsection{Results for the EFT model}
\label{subsec:resultsEFT}

In this section we present the combined constraints on the effective Higgs-portal setup for a \emph{vector} DM particle \(V\) with mass \(m_V\) and portal coupling \(\lambda_{HV}\). Unless stated otherwise, we assume that \(V\) accounts for the full DM abundance, \(\Omega_V=\Omega_{\rm DM}\). We combine the following constraints:
\begin{itemize}
\item \textbf{Cosmology:} the relic abundance must reproduce the observed value, \(\Omega_{\rm DM}h^2\simeq 0.120\) \cite{Planck:2018vyg}.
\item \textbf{DD:} limits on the SI DM--nucleon scattering cross section from LZ \cite{LZ:2024zvo} and the projected sensitivity from DARWIN \cite{DARWIN:2016hyl}.
\item \textbf{ID:} bounds on \(\langle\sigma v\rangle\) from the combined \(\gamma\)-ray analysis of dSphs with {\it Fermi}-LAT data \cite{McDaniel:2023bju}.
\item \textbf{Colliders:} the ATLAS upper limit on the invisible Higgs branching ratio, relevant for \(m_V<m_h/2\) \cite{ATLAS:2022yvh,Baek:2021hnl}.
\end{itemize}

\begin{figure}[t]
\centering
\includegraphics[width=0.99\linewidth]{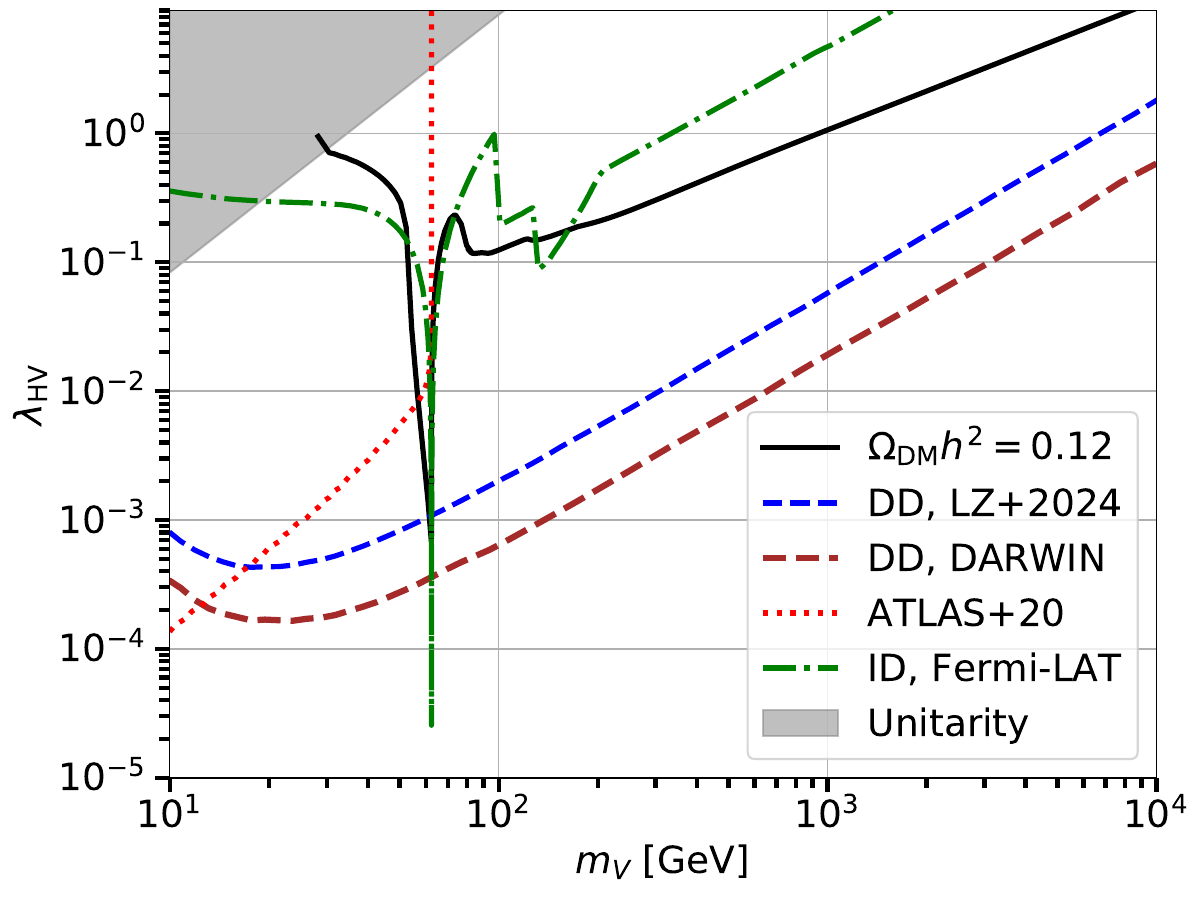}
\includegraphics[width=0.99\linewidth]{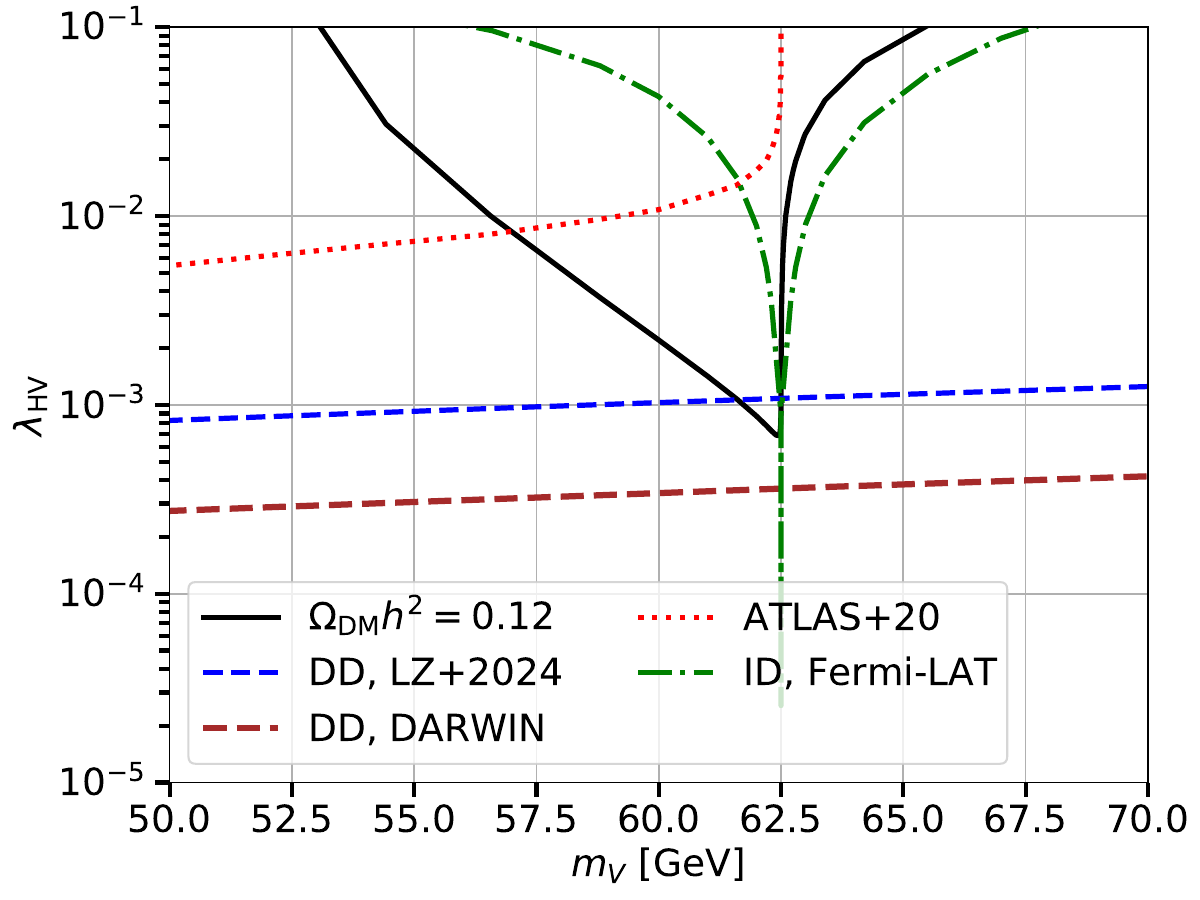}
\caption{\textbf{Combined constraints for the Higgs-portal EFT with vector DM \(V\).} The black solid curve shows the model parameters that satisfy the relic-density condition. The blue dashed curve shows the LZ upper limit on the SI scattering cross section recast as a bound on \(\lambda_{HV}\) \cite{LZ:2024zvo}, the brown dashed curve shows the projected DARWIN sensitivity recast in the same plane \cite{DARWIN:2016hyl}, the red dotted curve shows the ATLAS bound from invisible Higgs decays \(h\to VV\) (relevant only for \(m_V<m_h/2\)) \cite{ATLAS:2022yvh}, and the green dot-dashed curve shows the dSph \(\gamma\)-ray limits recast as bounds on \(\lambda_{HV}\) \cite{McDaniel:2023bju}. The gray shaded region indicates parameter values excluded by the perturbative-unitarity requirement discussed in Sec.~\ref{sec:validity}. The lower panel shows a zoom into the resonant region \(m_V\simeq m_h/2\), where the relic-density requirement can still be compatible with present direct-detection limits.}
\label{fig:resultsEFT}
\end{figure}

Figure~\ref{fig:resultsEFT} summarizes the resulting constraints in the \((m_V,\lambda_{HV})\) plane. The black solid curve shows the values of \(\lambda_{HV}\) that reproduce the observed relic density through thermal freeze-out. Away from the Higgs resonance, \(m_V\simeq m_h/2\), the correct abundance typically requires an electroweak-size portal coupling, \(\lambda_{HV}\sim \mathcal{O}(10^{-1})\) for \(m_V\) around the weak scale, as expected in a standard WIMP scenario. As \(m_V\) approaches the resonant region, the annihilation rate is enhanced by the \(s\)-channel Higgs propagator and the coupling required to obtain \(\Omega_V h^2\simeq 0.120\) drops sharply. Very close to the resonance one typically finds \(\lambda_{HV}\sim \mathcal{O}(10^{-3})\), as shown in the zoomed panel.

The ID bound (green dot-dashed) exhibits a qualitatively similar dip near \(m_V\simeq m_h/2\) when translated into an upper limit on \(\lambda_{HV}\). This behaviour follows directly from the Higgs propagator structure already implicit in Eq.~\eqref{eq:sigmavv_general_width}. Near the pole, the annihilation rate into a given SM final state \(X\) follows a Breit--Wigner form,
\begin{equation}
\sigma v \;\propto\;
\frac{\lambda_{HV}^2\,\Gamma_{h^\ast\to X}(s)}
{\bigl(s-m_h^2\bigr)^2+m_h^2\bigl[\Gamma_h^{\rm tot}(s)\bigr]^2}\,,
\,\, s\simeq 4m_V^2,
\label{eq:BW_V}
\end{equation}
up to kinematic factors. In dwarf spheroidals the DM velocities are tiny, \(v/c\sim 10^{-5}\), implying \(s=4m_V^2(1+\mathcal{O}(v^2))\) with a negligible shift in the center-of-mass energy. Therefore, ID probes essentially the intrinsic Breit--Wigner line shape, which is approximately symmetric around \(m_V=m_h/2\).

By contrast, the relic-density condition involves thermal averaging at freeze-out,
\begin{equation}
\langle\sigma v\rangle(x)=
\frac{x^{3/2}}{2\sqrt{\pi}}
\int_0^\infty dv\,v^2\,e^{-x v^2/4}\,\sigma v,
\qquad x\equiv \frac{m_V}{T},
\label{eq:thermalavg_V}
\end{equation}
with \(v_{\rm rms}\sim\sqrt{6/x_f}\approx 0.3\,c\) for \(x_f\sim 20\). For \(m_V<m_h/2\), part of the thermal velocity distribution can raise \(s\) enough to sample the resonance, while for \(m_V>m_h/2\) the distribution cannot move back toward the pole. This one-sided access to the resonance makes the relic-density curve around \(m_V\simeq m_h/2\) visibly asymmetric.

\medskip

The DD bounds exclude a large fraction of the parameter space because the SI scattering cross section scales as \(\sigma_{\rm SI}\propto \lambda_{HV}^2\) without any resonant enhancement. As a representative example, around \(m_V\simeq 100~{\rm GeV}\) the LZ limit implies \(\lambda_{HV}\lesssim \mathcal{O}(10^{-3})\) in this EFT setup, which is in strong tension with the freeze-out target away from the resonance. Consequently, the only region that remains compatible with both the relic density and current SI limits is a narrow strip around \(m_V\simeq m_h/2\), with \(\lambda_{HV}\sim \mathcal{O}(10^{-3})\). This region is expected to be thoroughly tested by next-generation experiments such as DARWIN.

Collider limits from invisible Higgs decays (red dotted) are relevant only for \(m_V<m_h/2\), where the decay \(h\to VV\) is kinematically open. In that mass range they provide a complementary constraint, but they become competitive with, or stronger than, DD only at sufficiently low masses where the nuclear-recoil sensitivity degrades. The gray shaded region labelled ``Unitarity'' indicates where the Proca EFT violates the perturbative-unitarity criterion discussed in Sec.~\ref{sec:validity}; this region should therefore not be regarded as a self-consistent EFT parameter space.

Overall, the Higgs-portal EFT remains viable only in a very narrow region close to the Higgs resonance, \(m_V\simeq m_h/2\). A convenient measure of the required proximity to resonance is \(\Delta\equiv |2m_V-m_h|/m_h\), which from Fig.~\ref{fig:resultsEFT} is at the per-mille level for points that satisfy both the relic-density requirement and current DD bounds. Such a strong localization around the resonance suggests an arguably unnatural degree of fine tuning in the DM mass, and this remaining parameter space is expected to be decisively tested by DARWIN.

\section{UV-complete model: gauged $U(1)_X$ realization with vector dark matter}
\label{sec:UV}

We consider a renormalizable UV completion of the vector Higgs-portal scenario based on a dark Abelian gauge symmetry $U(1)_X$.
The dark sector contains a massive vector field $V_\mu$ (the DM candidate) and a complex scalar $S$ charged under $U(1)_X$.
No SM field carries $U(1)_X$ charge.
Following the standard construction, a $\mathbb{Z}_2$ symmetry is imposed to forbid hypercharge--dark kinetic mixing and ensure the stability of $V_\mu$:
\begin{equation}
V_\mu \to -V_\mu,\,\,\, S \to S^\ast,\,\,\, \text{SM fields are $\mathbb{Z}_2$-even.}
\end{equation}
This discrete symmetry acts as a dark charge-conjugation symmetry and forbids the kinetic mixing operator $B_{\mu\nu}V^{\mu\nu}$.

\subsection{Lagrangian and scalar potential}

The Lagrangian is
\begin{equation}
\mathcal{L}=\mathcal{L}_{\rm SM}-\frac14\,V_{\mu\nu}V^{\mu\nu}+\left|D_\mu S\right|^2-V(\Phi,S)\,,
\end{equation}
where $V_{\mu\nu}=\partial_\mu V_\nu-\partial_\nu V_\mu$ and the covariant derivative acting on $S$ is
\begin{equation}
D_\mu S=\left(\partial_\mu-i g_X V_\mu\right)S\,,
\end{equation}
with $g_X$ the $U(1)_X$ gauge coupling.
We take the $U(1)_X$ charge of $S$ to be $+1$.

We adopt the following normalization for the tree-level scalar potential:
\begin{equation}
\label{eq:VHS}
V(\Phi,S) \!=\! -\frac{\mu_H^2}{2}\,|\Phi|^2-\frac{\mu_S^2}{2}\,|S|^2+\frac{\lambda_H}{4}|\Phi|^4+\frac{\lambda_S}{4}|S|^4+\frac{\lambda_{HS}}{4}|\Phi|^2|S|^2,
\end{equation}
where $\Phi$ is the SM Higgs doublet and $S$ is the dark Higgs.
The potential is bounded from below if
\begin{equation}
\label{eq:stability}
\lambda_H>0,\qquad \lambda_S>0,\qquad \lambda_{HS}>-2\sqrt{\lambda_H\lambda_S}\,.
\end{equation}

\subsection{Spontaneous symmetry breaking and the vector mass}

After electroweak and $U(1)_X$ symmetry breaking, we expand around the vacuum expectation values $v$ and $\omega$
\begin{equation}
\Phi(x)=\frac{1}{\sqrt{2}}
\begin{pmatrix}
0\\ v+h(x)
\end{pmatrix},
\,\,
S(x)=\frac{1}{\sqrt{2}}\left(\omega+s(x)+i\chi(x)\right),
\end{equation}
with $v\simeq 246~{\rm GeV}$ and $\omega>0$.
Here $h$ and $s$ denote the scalar fluctuations in the gauge basis.
In unitary gauge the Goldstone mode is removed, $\chi=0$.

From the kinetic term $|D_\mu S|^2$ one obtains
\begin{eqnarray}
&&|D_\mu S|^2 \supset \frac12\,g_X^2\,V_\mu V^\mu\,(\omega+s)^2 \\
&=& \frac12\,m_V^2\,V_\mu V^\mu+\left(\frac{m_V^2}{\omega}\right)s\,V_\mu V^\mu+\frac12\left(\frac{m_V^2}{\omega^2}\right)s^2\,V_\mu V^\mu\,,\nonumber
\end{eqnarray}
so that the DM mass is
\begin{equation}
\label{eq:mV}
m_V=g_X\,\omega\,.
\end{equation}
The remaining terms describe the interactions of the dark scalar fluctuation $s$ with the vector DM field.

\subsection{Minimization conditions and vevs}

Because the portal coupling $\lambda_{HS}$ mixes the two scalar sectors, the gauge-basis fields $h$ and $s$ are not mass eigenstates after symmetry breaking.

At the minimum, the tadpole conditions $\partial V/\partial h=0$ and $\partial V/\partial s=0$ yield
\begin{equation}
\label{eq:tadpoles}
\mu_H^2=\frac{\lambda_H}{2}\,v^2+\frac{\lambda_{HS}}{4}\,\omega^2,
\,\,
\mu_S^2=\frac{\lambda_S}{2}\,\omega^2+\frac{\lambda_{HS}}{4}\,v^2.
\end{equation}
Solving for the vevs in terms of the Lagrangian parameters gives
\begin{equation}
\label{eq:vev_solutions}
v^2=\frac{8\mu_H^2\lambda_S-4\lambda_{HS}\mu_S^2}{4\lambda_H\lambda_S-\lambda_{HS}^2},
\,\,
\omega^2=\frac{8\mu_S^2\lambda_H-4\lambda_{HS}\mu_H^2}{4\lambda_H\lambda_S-\lambda_{HS}^2}.
\end{equation}
These expressions are the translation of the standard Higgs-portal results into the normalization of Eq.~\eqref{eq:VHS}.

\subsection{Scalar mass matrix, eigenstates, and mixing}

Expanding the potential to quadratic order in $(h,s)$ and using the tadpole relations in Eq.~\eqref{eq:tadpoles}, the scalar mass terms can be written as
\begin{equation}
\label{eq:mass_terms}
V \supset \frac12\,(h\ \ s)
\begin{pmatrix}
M_{hh}^2 & M_{hs}^2\\
M_{hs}^2 & M_{ss}^2
\end{pmatrix}
\binom{h}{s},
\end{equation}
with
\begin{equation}
\label{eq:mass_matrix_entries}
M_{hh}^2=\frac12\,\lambda_H v^2,
\,\,
M_{ss}^2=\frac12\,\lambda_S \omega^2,
\,\,
M_{hs}^2=\frac14\,\lambda_{HS}\,v\,\omega.
\end{equation}

We define the mass eigenstates $(H_1,H_2)$ through the orthogonal rotation
\begin{equation}
\binom{h}{s}=
\begin{pmatrix}
\cos\theta & -\sin\theta\\
\sin\theta & \cos\theta
\end{pmatrix}
\binom{H_1}{H_2},
\end{equation}
where $H_1$ is identified with the observed SM Higgs boson, $m_{H_1}\simeq125~{\rm GeV}$.
The mixing angle satisfies
\begin{equation}
\label{eq:tan2theta}
\tan(2\theta)=\frac{2M_{hs}^2}{M_{hh}^2-M_{ss}^2}
=\frac{\lambda_{HS}\,v\,\omega}{\lambda_H v^2-\lambda_S \omega^2}.
\end{equation}
The physical masses are
\begin{equation}
\label{eq:masses}
m_{H_{1,2}}^2\!\!=
\frac14\left(\lambda_H v^2+\lambda_S\omega^2
\mp
\sqrt{(\lambda_H v^2-\lambda_S\omega^2)^2+(\lambda_{HS}v\omega)^2}\right).
\end{equation}

\subsection{Useful reparameterization in terms of physical inputs}

For phenomenology it is often convenient to trade the Lagrangian parameters for a set of physical inputs.
After fixing $m_{H_1}$ and $v$ fixed by SM measurements ($m_{H_1}\simeq125~{\rm GeV}$ and $v=246~{\rm GeV}$) a practical choice is
\begin{equation}
\{m_{H_2},\,\theta,\,\lambda_{HS},\,m_V\}.
\end{equation}

The dark vev $\omega$ is obtained from the off-diagonal element of the scalar mass matrix:
\begin{equation}
\label{eq:omega_from_inputs_relation}
M_{hs}^2=\frac12\left(m_{H_2}^2-m_{H_1}^2\right)\sin(2\theta)=\frac14\,\lambda_{HS}\,v\,\omega\,,
\end{equation}
which gives
\begin{equation}
\label{eq:omega_from_inputs}
\omega=\frac{2\left(m_{H_2}^2-m_{H_1}^2\right)\sin(2\theta)}{\lambda_{HS}\,v}\,.
\end{equation}
The quartic couplings then follow from $M=R\,{\rm diag}(m_{H_1}^2,m_{H_2}^2)\,R^T$:
\begin{align}
\label{eq:lambdas_from_inputs}
\lambda_H &= \frac{2}{v^2}\left(m_{H_1}^2\cos^2\theta+m_{H_2}^2\sin^2\theta\right),\\
\lambda_S &= \frac{2}{\omega^2}\left(m_{H_1}^2\sin^2\theta+m_{H_2}^2\cos^2\theta\right).
\end{align}
Finally, the gauge coupling is fixed by the vector mass,
\begin{equation}
\label{eq:gx_from_inputs}
g_X=\frac{m_V}{\omega}.
\end{equation}
If needed, the quadratic parameters $\mu_H^2$ and $\mu_S^2$ can then be reconstructed from Eq.~\eqref{eq:tadpoles}.

\subsection{Couplings relevant for DM phenomenology}

Because only the $h$ component couples directly to SM states, the couplings of $H_1$ and $H_2$ to SM fermions and gauge bosons are universally rescaled:
\begin{equation}
g_{H_1{\rm SM}}=\cos\theta\;g_{h{\rm SM}}^{\rm SM},
\qquad
g_{H_2{\rm SM}}=\sin\theta\;g_{h{\rm SM}}^{\rm SM}.
\end{equation}
On the other hand, the dark vector couples to the $s$ field.
Using the interaction terms from $|D_\mu S|^2$, one finds
\begin{equation}
\mathcal{L}\supset \left(\frac{m_V^2}{\omega}\right)\left(\sin\theta\,H_1+\cos\theta\,H_2\right)V_\mu V^\mu.
\end{equation}
Equivalently, if one writes the interactions as
\begin{equation}
\mathcal{L}\supset \frac12\,g_{H_1VV}\,H_1\,V_\mu V^\mu+\frac12\,g_{H_2VV}\,H_2\,V_\mu V^\mu,
\end{equation}
the couplings are
\begin{eqnarray}
g_{H_1VV}=2\,\frac{m_V^2}{\omega}\sin\theta=2\,g_X\,m_V\,\sin\theta, \\
g_{H_2VV}=2\,\frac{m_V^2}{\omega}\cos\theta=2\,g_X\,m_V\,\cos\theta.
\end{eqnarray}
Therefore, in the limit of small mixing $\sin(\theta)\ll 1$, $H_1$ couples mostly to SM fermions while $H_2$ to DM.

\subsection{Direct Detection (UV-complete model)}

Elastic scattering on nucleons is mediated by $t$-channel exchange of $H_1$ and $H_2$.
At leading order, the SI DM--nucleon cross section is given by \cite{Ko:2014gha}
\begin{equation}
\label{eq:DD_UV}
\sigma_{\rm SI}^{(N)}\!\!=\!
\frac{\mu_{VN}^2}{4\pi}\!
\left(\frac{m_N f_N}{v}\right)^2\!\!
g_X^2\,\sin^2(2\theta)\!
\left(\frac{1}{m_{H_1}^2}-\frac{1}{m_{H_2}^2}\right)^2,
\end{equation}
where $\mu_{VN}=m_V m_N/(m_V+m_N)$ is the reduced mass and $f_N$ parameterizes the scalar nucleon matrix element.
This expression makes explicit the characteristic interference 
between the $H_1$- and $H_2$-exchange amplitudes. In 
particular, the cross section vanishes for $\theta \rightarrow 0$, 
as expected when the visible and dark sectors decouple. In a 
genuine decoupling regime with $m_{H_2} \rightarrow \infty$, 
the mixing angle also tends to zero for fixed underlying 
couplings, so that the full cross section becomes suppressed.

It is instructive to verify explicitly that the EFT result of 
Sec.~\ref{sec:eft} is recovered in the appropriate decoupling 
limit. In the limit $m_{H_2} \gg m_{H_1}$ with fixed 
$\sin\theta \simeq \theta \ll 1$, the mixing angle scales as
\begin{equation}
    \sin\theta \simeq \frac{\lambda_{HS}\, v\,\omega}{m_{H_2}^2 
    - m_{H_1}^2} \approx \frac{\lambda_{HS}\, v\,\omega}{m_{H_2}^2},
\end{equation}
and the interference factor in Eq.~\eqref{eq:DD_UV} reduces to
\begin{equation}
    g_X^2\sin^2(2\theta)\left(\frac{1}{m_{H_1}^2} 
    - \frac{1}{m_{H_2}^2}\right)^2 
    \xrightarrow{m_{H_2}\to\infty}\ 
    \frac{\lambda_{HV}^2\, v^2}{m_{H_1}^4},
\end{equation}
where we have identified the effective EFT coupling via 
$\lambda_{HV}\,v \equiv g_{H_1 VV} = 2g_X m_V \sin\theta$. 
Substituting into Eq.~\eqref{eq:DD_UV}, one recovers
\begin{equation}
    \sigma^{(N)}_\mathrm{SI} 
    \xrightarrow{m_{H_2}\to\infty}\ 
    \frac{\lambda_{HV}^2\, f_N^2}{4\pi\, m_{H_1}^4}
    \frac{\mu_{VN}^2\, m_N^2}{m_V^2},
    \label{eq:EFT_limit}
\end{equation}
in agreement with Eq.~\eqref{eq:sigma_eft_final}. This confirms that 
the EFT expression is the leading term in a $1/m_{H_2}^2$ 
expansion of the UV-complete model, and that the additional 
suppression in the full model relative to the naive EFT estimate 
arises from the destructive interference between the two 
scalar-exchange amplitudes when $m_{H_2}$ is finite.
\subsection{Invisible and exotic decays of the SM-like Higgs}

If $m_V<m_{H_1}/2$, the SM-like Higgs $H_1$ can decay invisibly into a pair of DM vectors.
The partial width is \cite{Ko:2014loa}
\begin{eqnarray}
&&\Gamma(H_1\to VV)=
\frac{g_{H_1VV}^{\,2}\,m_{H_1}^3}{128\pi\,m_V^4} \cdot \nonumber\\
&\cdot&\left(1-\frac{4m_V^2}{m_{H_1}^2}+\frac{12 m_V^4}{m_{H_1}^4}\right)
\sqrt{1-\frac{4m_V^2}{m_{H_1}^2}}.
\label{eq:Gamma_H1_VV}
\end{eqnarray}
Using the couplings above, this width is fully determined by the set \(\{m_{H_1},m_V,\sin\theta,\omega\}\), or equivalently by the chosen physical input parameters of the model.

If $m_{H_2}<m_{H_1}/2$, the decay into two lighter scalars is kinematically allowed.
Defining the trilinear coupling through
\begin{equation}
\mathcal{L}\supset -\frac12\,g_{122}\,H_1H_2H_2,
\end{equation}
the corresponding partial width reads
\begin{equation}
\Gamma(H_1\to H_2H_2)=
\frac{g_{122}^{\,2}}{32\pi\,m_{H_1}}\,
\sqrt{1-\frac{4m_{H_2}^2}{m_{H_1}^2}}.
\label{eq:Gamma_H1_H2H2_general}
\end{equation}
Here $g_{122}$ is the cubic scalar coupling obtained by expanding the potential in Eq.~\eqref{eq:VHS} to third order and rotating to the $(H_1,H_2)$ basis.
Its explicit form depends on the normalization convention adopted for the scalar potential, so it should be derived consistently within the conventions used in this section.

While \(\Gamma(H_1\to VV)\) contributes directly to the invisible Higgs width, the decay \(H_1\to H_2H_2\) is experimentally invisible only if \(H_2\) decays dominantly into dark states or is sufficiently long-lived.
Otherwise it contributes to an exotic visible Higgs decay.

The total width of \(H_1\) can therefore be written as
\begin{widetext}
\begin{equation}
\Gamma_{H_1}^{\rm tot}=
\cos^2\theta\,\Gamma_{h}^{\rm SM}(m_{H_1})
+\Gamma(H_1\to VV)\,\Theta(m_{H_1}-2m_V)
+\Gamma(H_1\to H_2H_2)\,\Theta(m_{H_1}-2m_{H_2}),
\label{eq:GammaH1_tot}
\end{equation}
\end{widetext}
where \(\Gamma_{h}^{\rm SM}(m_{H_1})\) is the SM Higgs total width evaluated at \(m_{H_1}\simeq125~\mathrm{GeV}\), namely \(\Gamma_{h}^{\rm SM}\simeq 4.07~\mathrm{MeV}\).

If one is interested specifically in the branching ratio into vector DM, one should use
\begin{equation}
{\rm Br}(H_1\to VV)=
\frac{\Gamma(H_1\to VV)\,\Theta(m_{H_1}-2m_V)}
{\Gamma_{H_1}^{\rm tot}}.
\label{eq:BR_H1_to_VV}
\end{equation}
If instead \(H_2H_2\) is also experimentally invisible, the quantity to compare with invisible-Higgs searches is
\begin{widetext}
\begin{equation}
{\rm Br}_{\rm inv}^{\rm (exp)}(H_1)=
\frac{\Gamma(H_1\to VV)\,\Theta(m_{H_1}-2m_V)+\Gamma(H_1\to H_2H_2)\,\Theta(m_{H_1}-2m_{H_2})}
{\Gamma_{H_1}^{\rm tot}}.
\label{eq:BRinv_exp_H1}
\end{equation}
\end{widetext}
This is the appropriate quantity to compare with experimental limits such as those reported in Ref.~\cite{ATLAS:2022yvh}.

\section{Results for the UV-complete model}
\label{sec:results}

\begin{figure*}[t]
\centering
\includegraphics[width=0.325\linewidth]{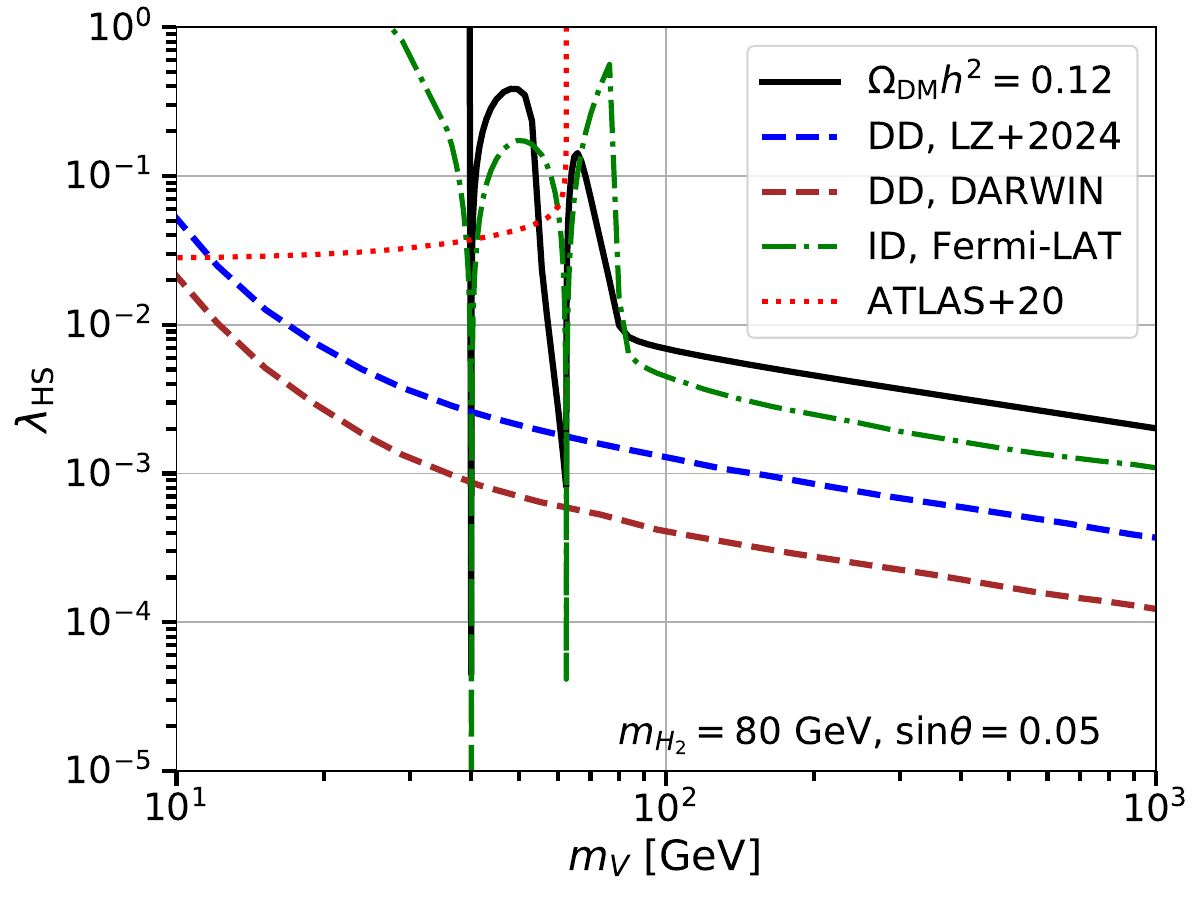}
\includegraphics[width=0.325\linewidth]{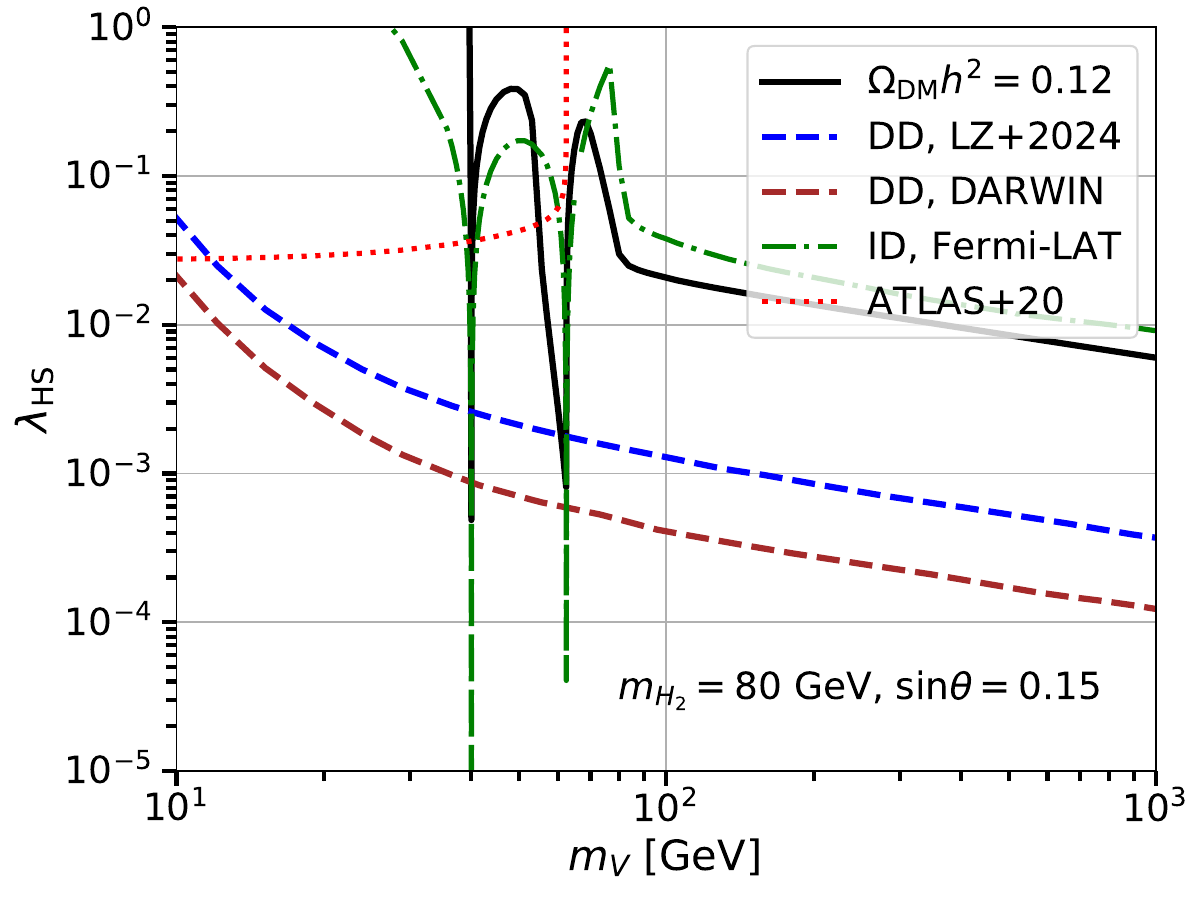}
\includegraphics[width=0.325\linewidth]{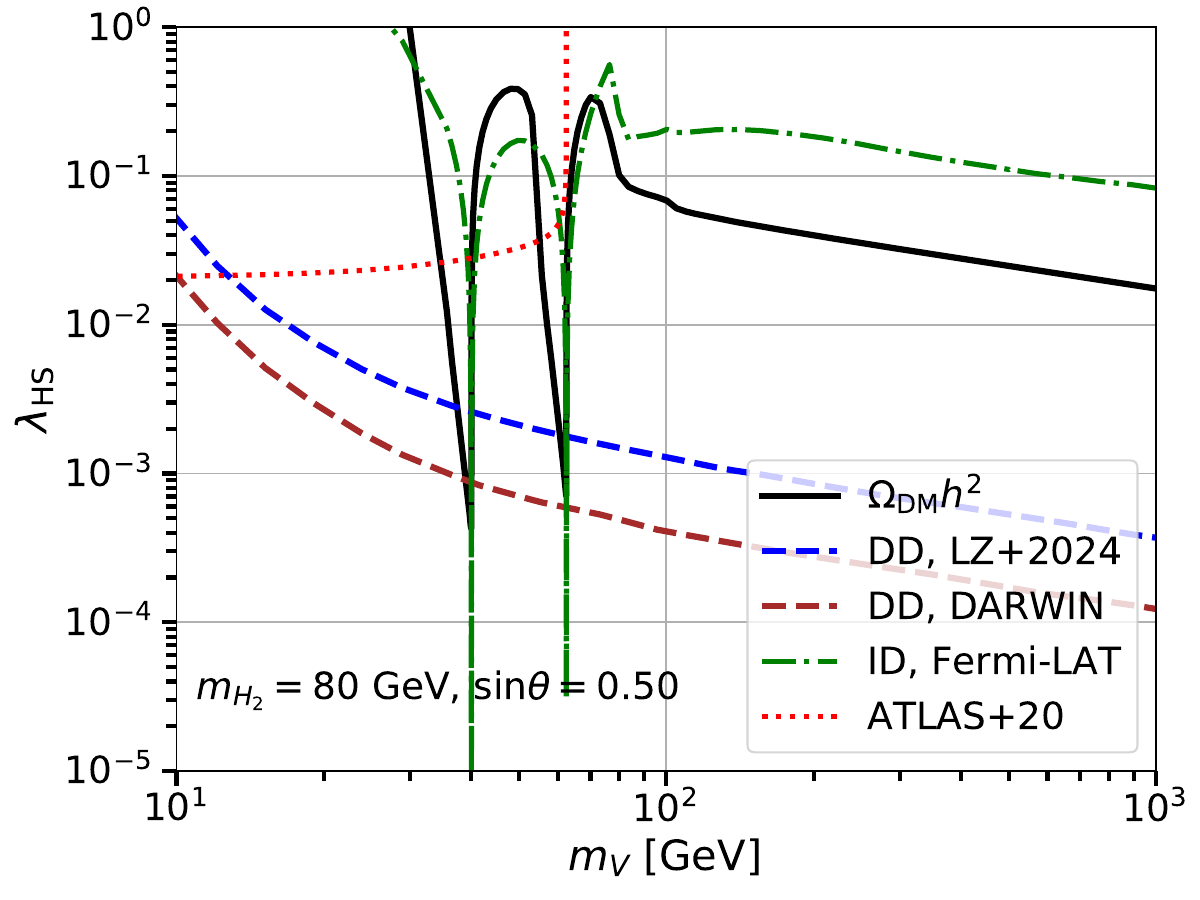}
\includegraphics[width=0.325\linewidth]{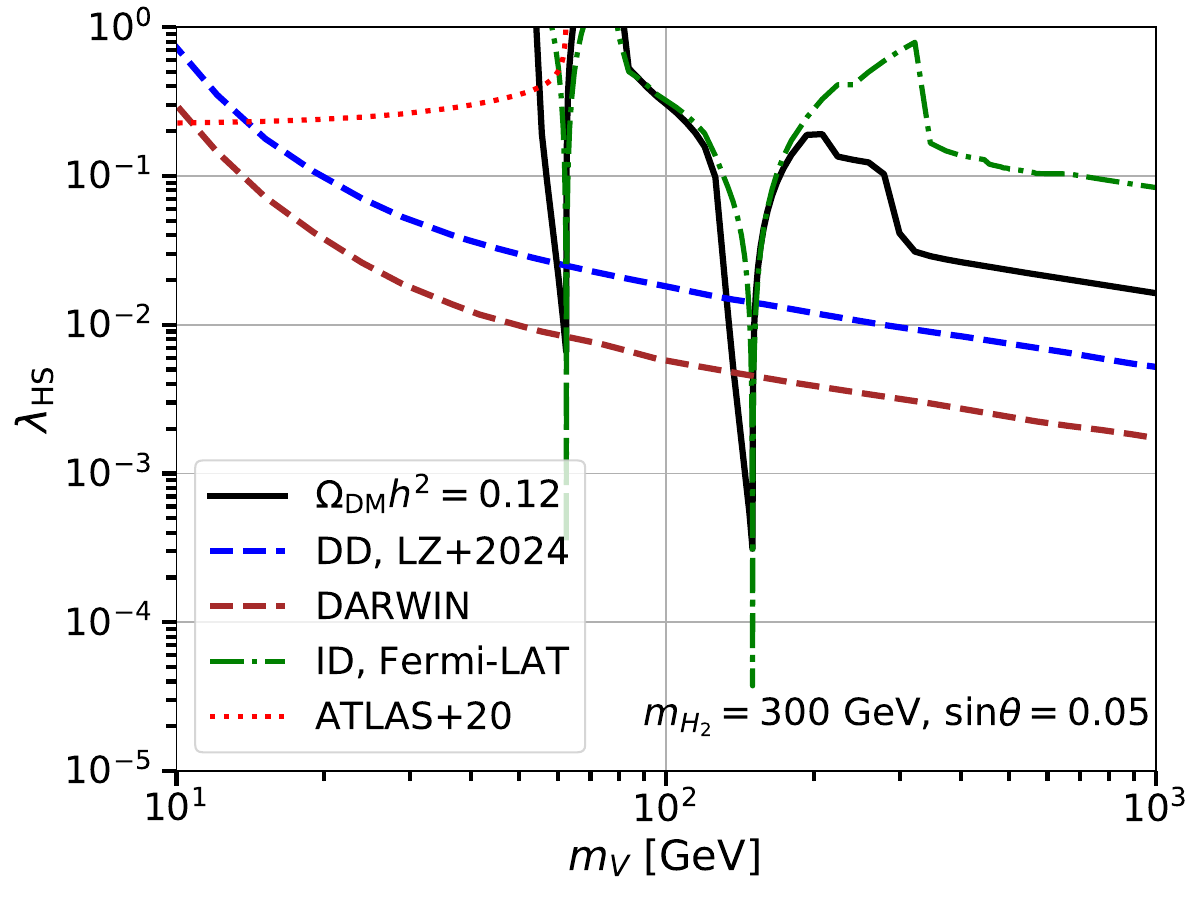}
\includegraphics[width=0.325\linewidth]{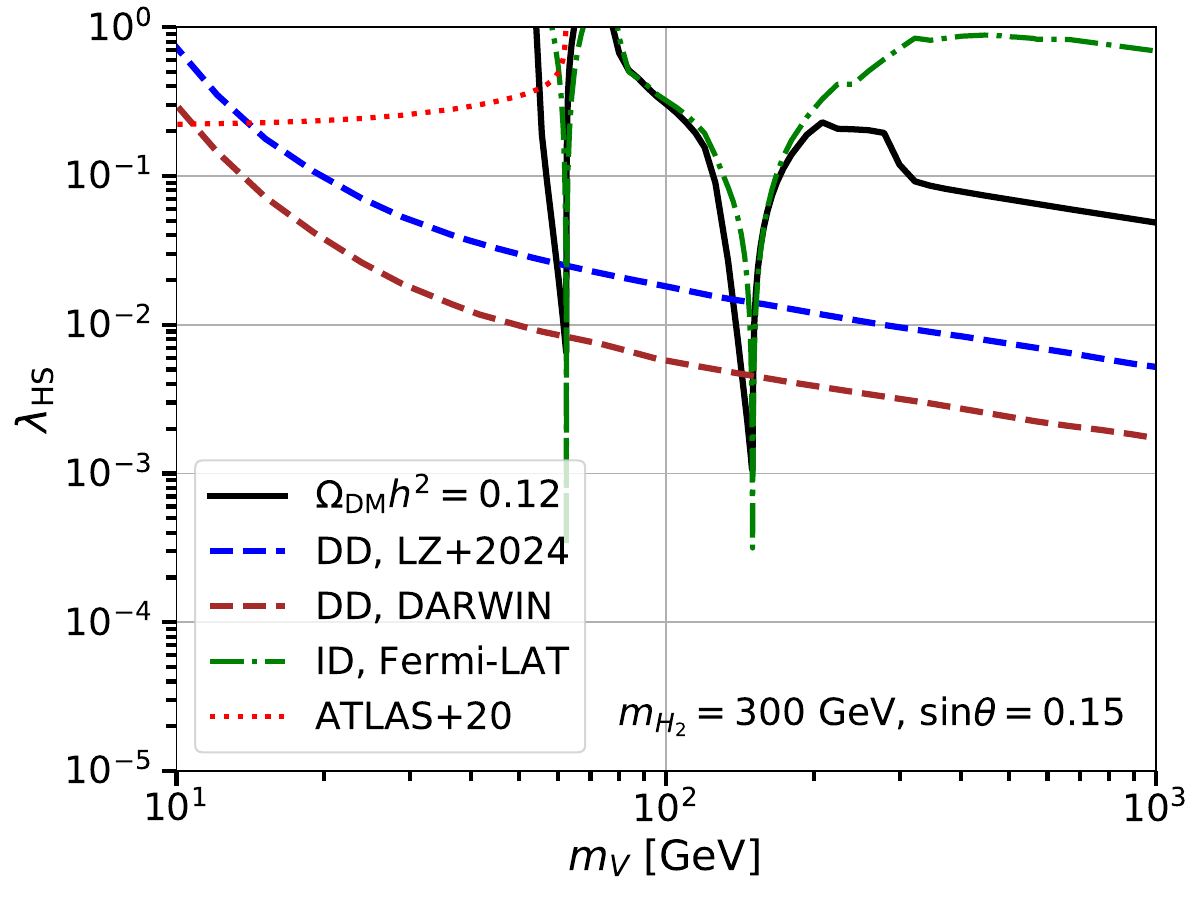}
\includegraphics[width=0.325\linewidth]{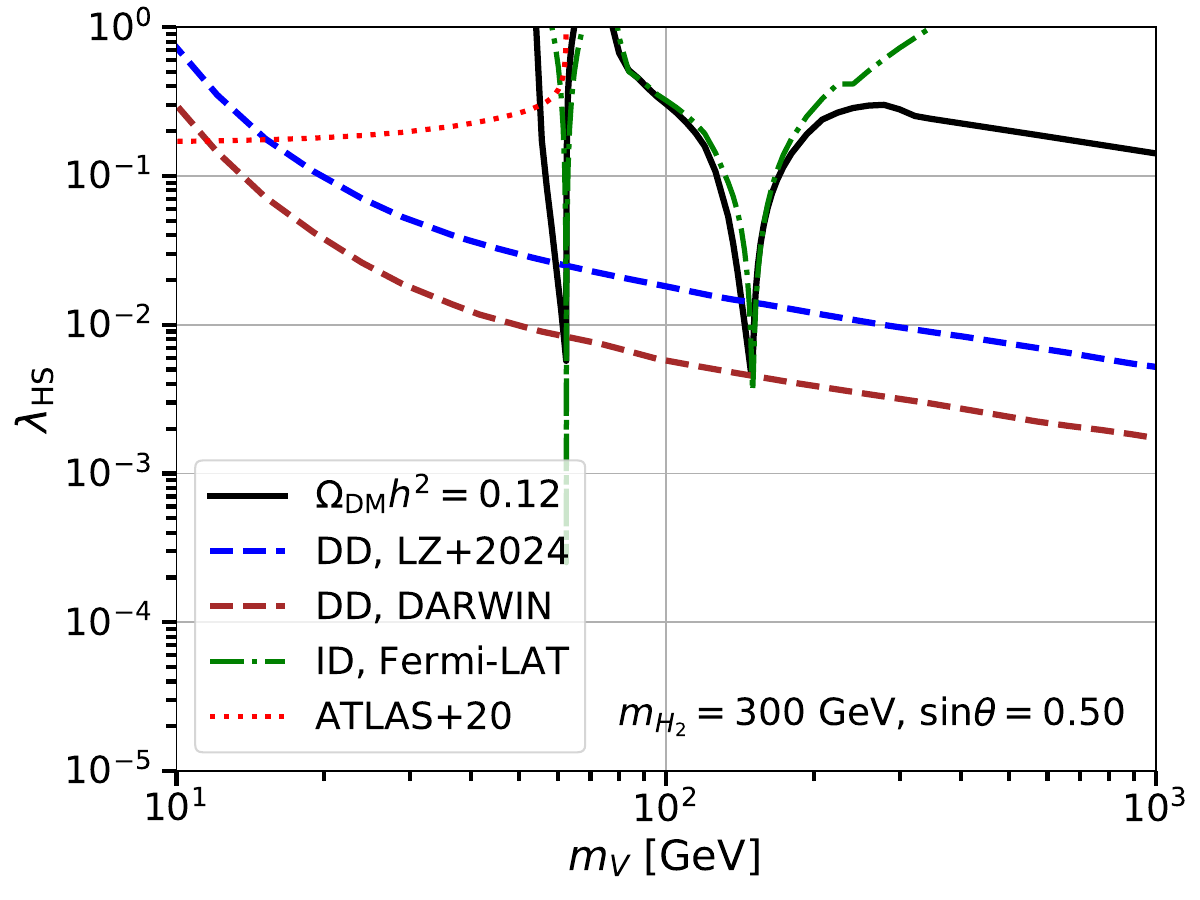}
\includegraphics[width=0.325\linewidth]{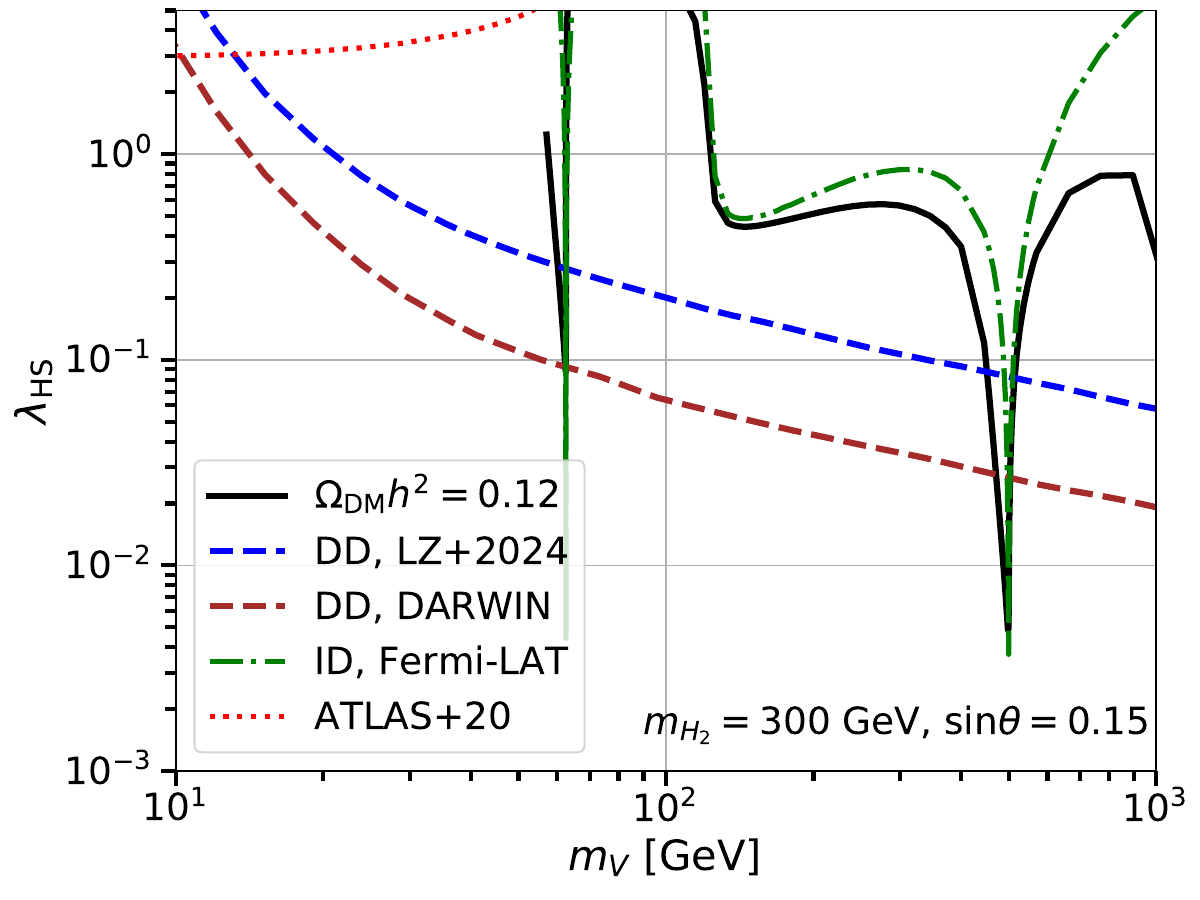}
\includegraphics[width=0.325\linewidth]{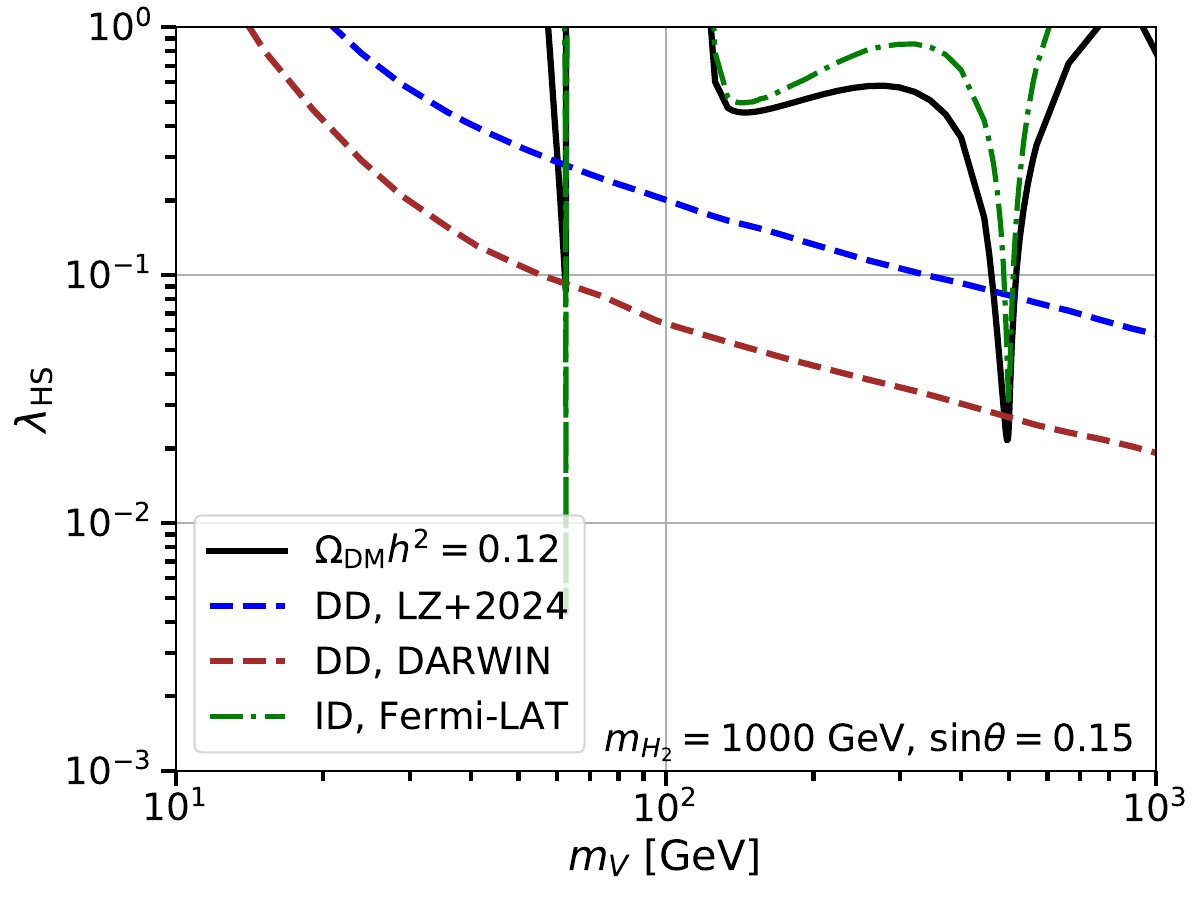}
\includegraphics[width=0.325\linewidth]{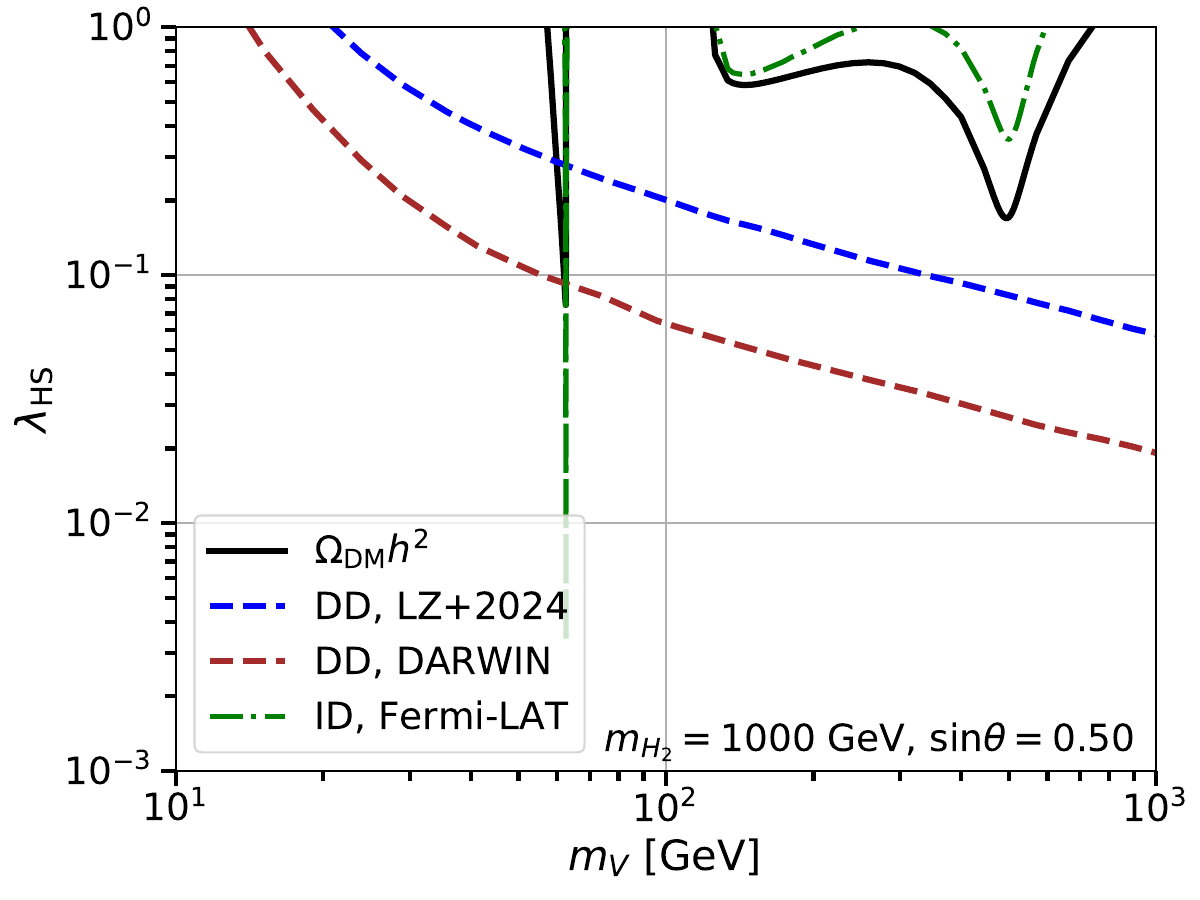}
\caption{\textbf{Combined constraints for the UV-complete vector Higgs-portal model.} For each benchmark choice of the heavy scalar mass \(m_{H_2}\) (rows) and mixing angle \(\sin\theta\) (columns), the black solid curve shows the value of \(\lambda_{HS}\) that reproduces the observed relic abundance \(\Omega_{\rm DM}h^2\simeq 0.12\). The blue dashed curve is the current LZ limit on the spin-independent DM--nucleon scattering cross section, recast as an upper bound on \(\lambda_{HS}\) \cite{LZ:2024zvo}. The red dashed curve shows the projected DARWIN sensitivity, also recast as a bound on \(\lambda_{HS}\) \cite{DARWIN:2016hyl}. The green dot-dashed curve shows the dwarf-spheroidal \(\gamma\)-ray limits from \textit{Fermi}-LAT, recast as bounds on \(\lambda_{HS}\) \cite{McDaniel:2023bju}. Where visible, the red dotted curve shows the ATLAS bound from invisible Higgs decays, relevant for \(m_V<m_{H_1}/2\) \cite{ATLAS:2022yvh}. In each panel, viable thermal-relic solutions require the relic-density curve to lie below the relevant experimental bounds.}
\label{fig:resultsUV}
\end{figure*}

Figure~\ref{fig:resultsUV} summarizes the interplay among relic density, DD, ID, and collider constraints in the UV-complete vector Higgs-portal model. We consider the benchmark choices \(m_{H_2}=(80,\,300,\,1000)\,\mathrm{GeV}\) and \(\sin\theta=(0.05,\,0.15,\,0.50)\), thus covering the cases of a second scalar lighter than, comparable to, and much heavier than the SM-like Higgs, together with small, intermediate, and large scalar mixing. In each panel, the black curve gives the value of \(\lambda_{HS}\) required to reproduce \(\Omega_{\rm DM}h^2\simeq 0.12\). The DD, ID, and collider curves are upper limits on \(\lambda_{HS}\): regions above them are excluded, while viable thermal-relic solutions require the black curve to lie below the strongest bound.

Two generic features appear throughout the figure. First, the relic-density curve exhibits dips near \(m_V\simeq m_{H_1}/2\) and \(m_V\simeq m_{H_2}/2\), corresponding to resonant annihilation through the two CP-even scalar mediators \(H_1\) and \(H_2\). Near these resonances, the annihilation cross section at freeze-out is enhanced, so that the observed relic abundance can be obtained with substantially smaller values of \(\lambda_{HS}\). Second, the same resonant structure is reflected in the ID limits, since the present-day annihilation rate is also enhanced when the mediator is close to the pole. As discussed for the EFT case, the ID limits are approximately symmetric around \(m_V=m_{H_i}/2\), while the relic-density curve is typically asymmetric because of thermal averaging at freeze-out.

Around the \(H_1\) resonance, the behaviour is qualitatively similar to that found in the EFT case shown in Fig.~\ref{fig:resultsEFT}. In particular, the correct relic abundance can be reproduced for \(m_V\simeq m_{H_1}/2\) with \(\lambda_{HS}\) in the approximate range \(10^{-3}\)–\(10^{-2}\), depending on the benchmark. Current DD bounds remove most of this region, leaving only the immediate vicinity of the resonance as potentially viable. For \(m_V<m_{H_1}/2\), invisible Higgs decays provide an additional constraint, shown by the red dotted curve where applicable.

The main qualitative difference with respect to the EFT description arises from the second scalar resonance at \(m_V\simeq m_{H_2}/2\). For small mixing, the coupling of \(H_2\) to SM states is suppressed, and the resulting resonance can efficiently deplete the DM abundance while keeping DD rates small. This behaviour is most evident for \(\sin\theta=0.05\), where the correct relic density can be achieved in a narrow window around \(m_V\simeq m_{H_2}/2\) while satisfying both current and projected DD bounds. For example, for \(m_{H_2}=300~\mathrm{GeV}\), the thermal-relic solution near \(m_V\simeq150~\mathrm{GeV}\) lies below the LZ and DARWIN limits, whereas away from resonance the coupling required by freeze-out increases rapidly and becomes excluded by DD. In these benchmarks, the ID limits are most constraining in the immediate vicinity of the resonance, while they are typically weaker than DD away from the pole.

\begin{figure*}[t]
\centering
\includegraphics[width=0.49\linewidth]{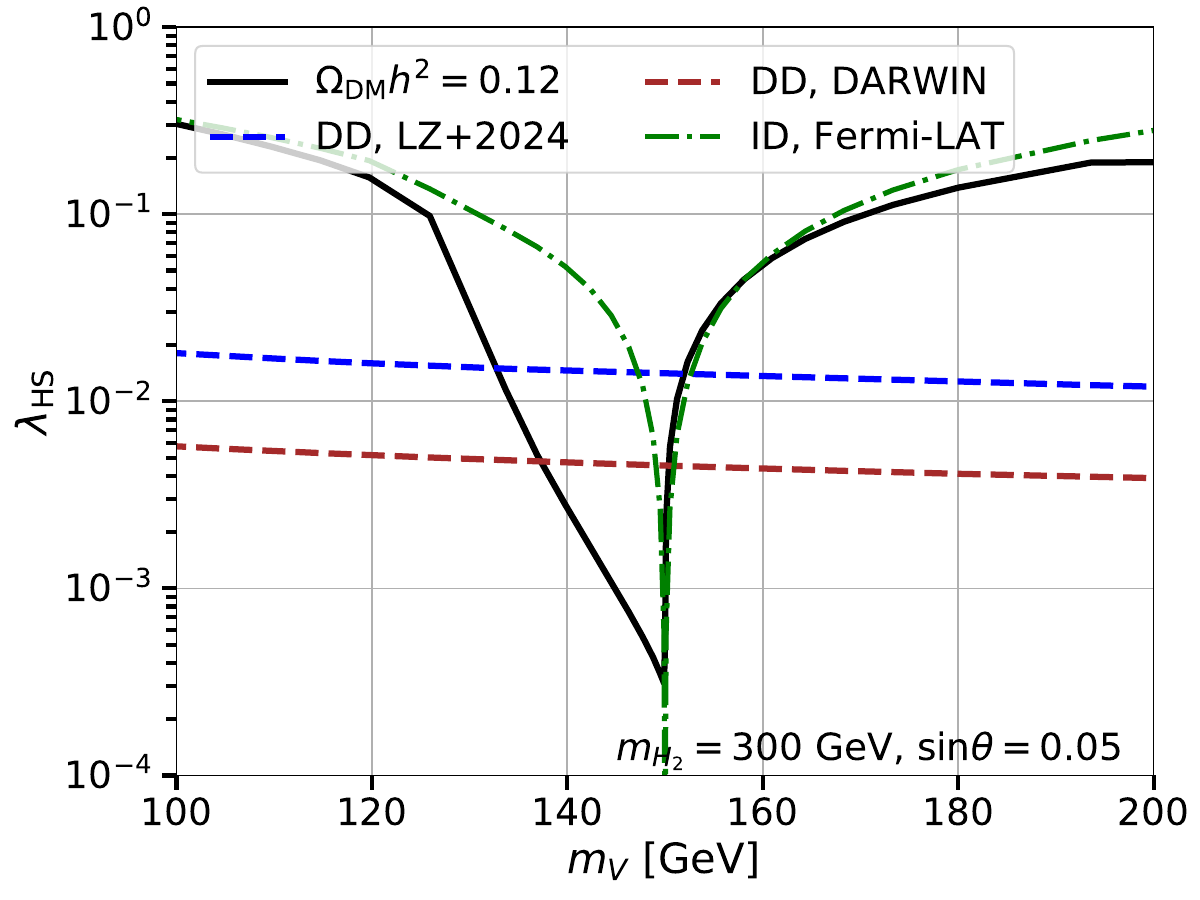}
\includegraphics[width=0.49\linewidth]{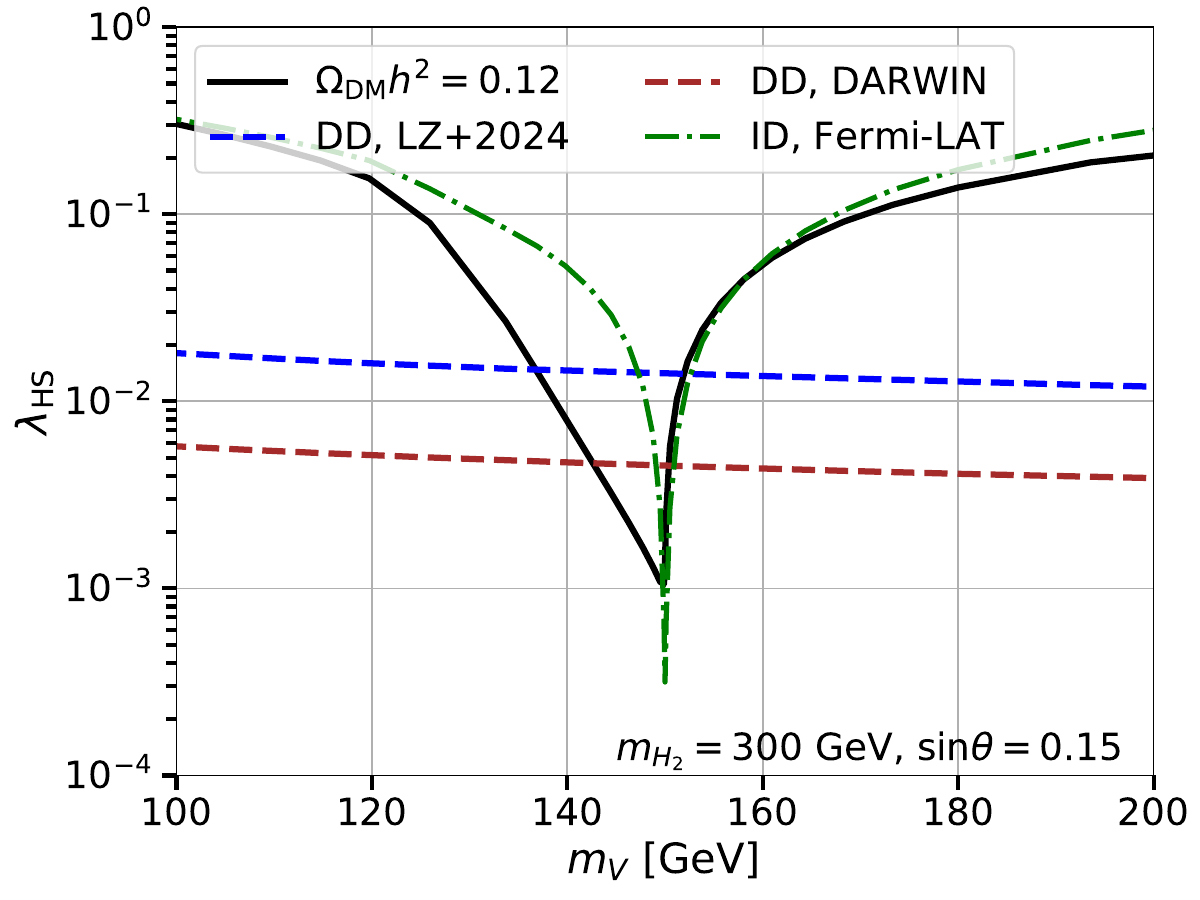}
\includegraphics[width=0.49\linewidth]{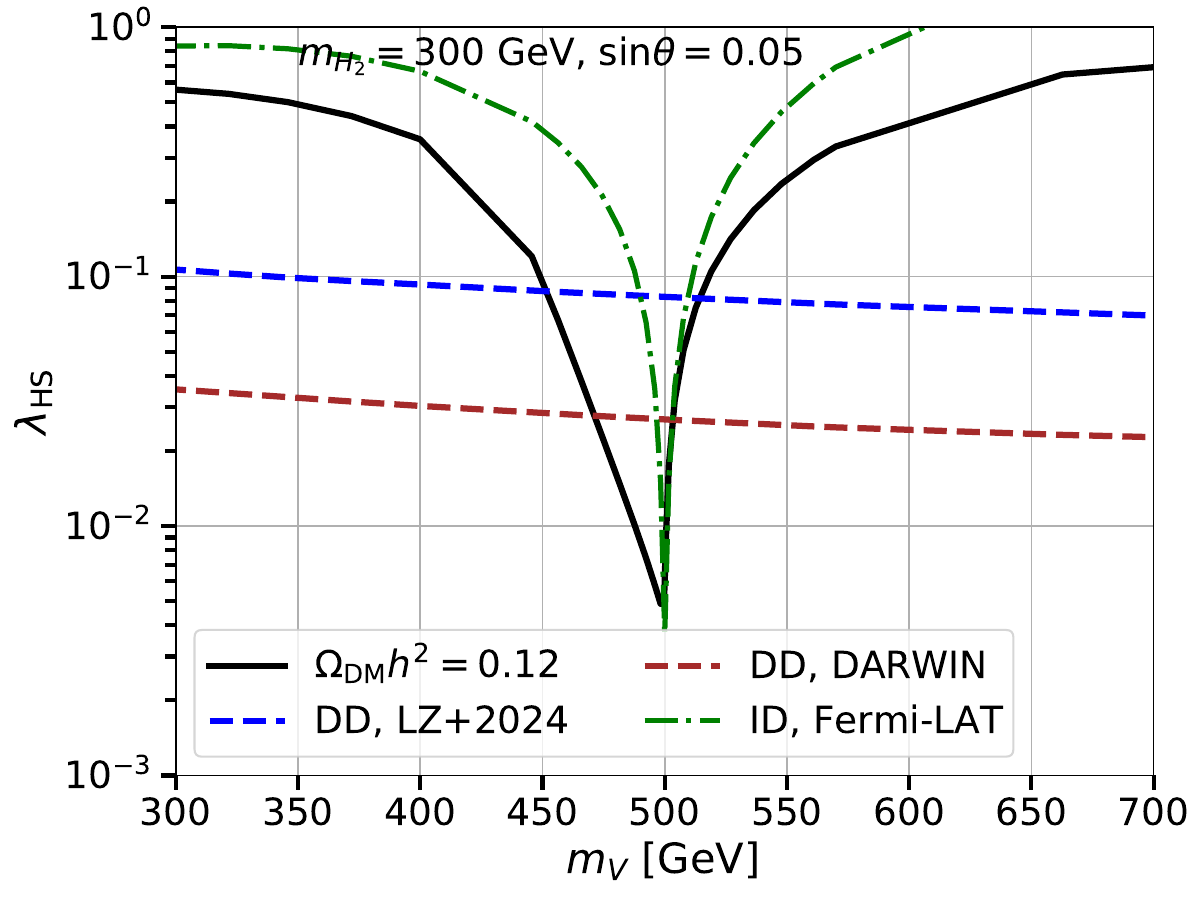}
\includegraphics[width=0.49\linewidth]{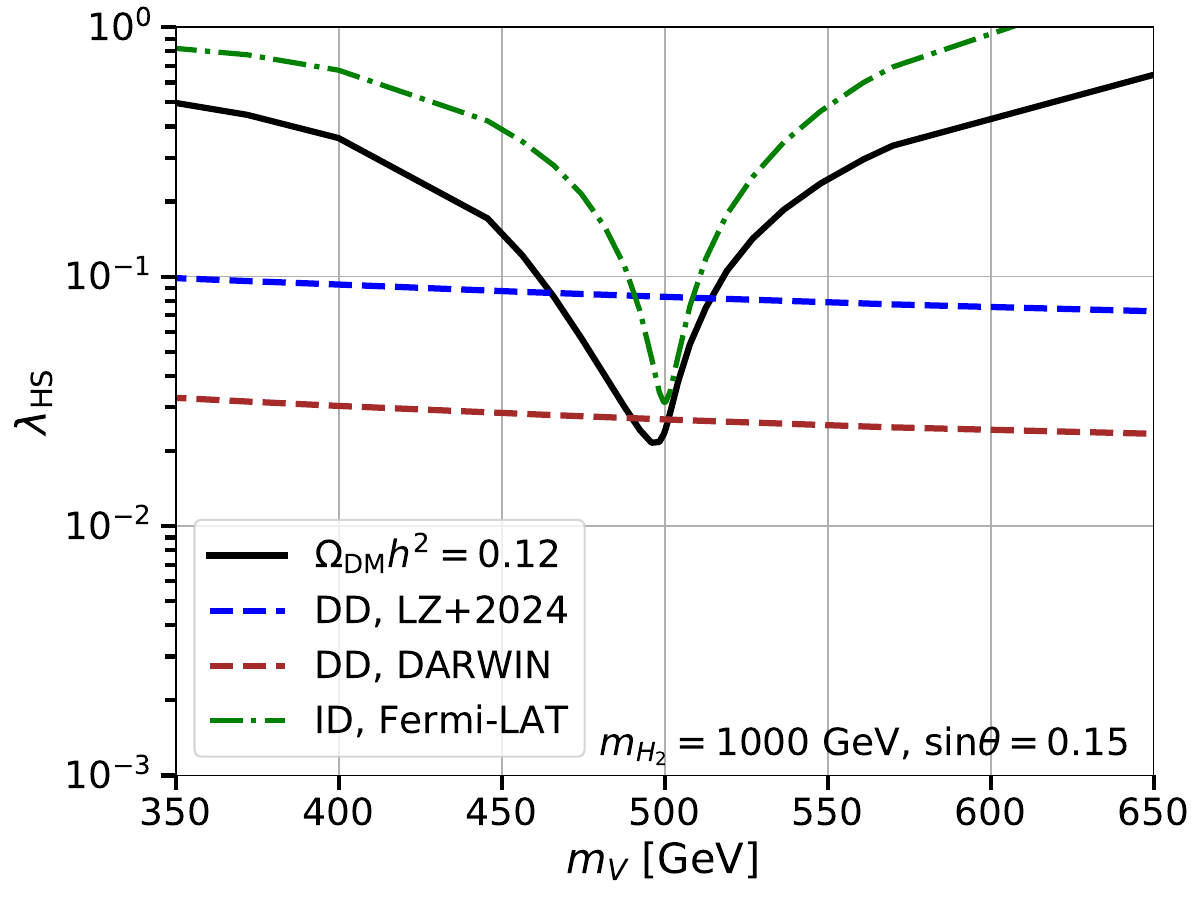}
\caption{\textbf{Zoom into the \(H_2\) resonance region.} Same quantities as in Fig.~\ref{fig:resultsUV}, but focusing on the resonant annihilation region \(m_V\simeq m_{H_2}/2\). The top row shows \(m_{H_2}=300~\mathrm{GeV}\) for \(\sin\theta=0.05\) (left) and \(\sin\theta=0.15\) (right), while the bottom row shows \(m_{H_2}=1000~\mathrm{GeV}\) for \(\sin\theta=0.05\) (left) and \(\sin\theta=0.15\) (right). Near the resonance, the correct relic density can be achieved with values of \(\lambda_{HS}\) significantly smaller than those required off resonance, thereby opening viable parameter space below present (LZ) and projected (DARWIN) DD limits.}
\label{fig:resultsUVzoom}
\end{figure*}

For larger mixing angles, \(\sin\theta=0.15\) and \(0.50\), the viable region generally shrinks and becomes more tightly localized around the resonant strips. This reflects the fact that increasing the mixing strengthens the coupling of the scalar mediators to SM states and typically enhances the DD sensitivity in the \((m_V,\lambda_{HS})\) plane. As a consequence, for \(\sin\theta>0.05\) the parameter space compatible with current DD constraints is reduced and pushed closer to the resonant region, as illustrated in Fig.~\ref{fig:resultsUVzoom}.

Overall, the UV completion provides a qualitative advantage over the Proca EFT description: the presence of a second scalar mediator opens an additional resonant annihilation channel, while DD can remain suppressed for sufficiently small mixing. The viable parameter space is therefore localized near \(m_V\simeq m_{H_2}/2\). A useful measure of the required proximity to resonance is \(\Delta \equiv |2m_V-m_{H_2}|/m_{H_2}\). From Fig.~\ref{fig:resultsUVzoom}, viable solutions typically require \(\Delta\) at the level of a few percent up to \(\mathcal{O}(10\%)\), depending on the benchmark point and on whether one imposes the current LZ bound or the projected DARWIN sensitivity.

Finally, the benchmark points shown in Fig.~\ref{fig:resultsUV} should also be interpreted in light of collider constraints on scalar mixing and on non-SM Higgs decays. In particular, for \(m_V<m_{H_1}/2\), the decay \(H_1\to VV\) contributes to the invisible Higgs width and can exclude part of the low-mass region even when DD is relatively weak. Moreover, large values of the mixing angle, such as \(\sin\theta=0.50\), are generally expected to be in tension with Higgs signal-strength measurements and direct searches for additional scalars. For this reason, the panels with large \(\sin\theta\) should mainly be regarded as illustrating the parametric dependence of the model, whereas the most robust viable regions are those with smaller mixing.

Overall, the UV-complete model remains viable in resonant regions that are absent or strongly reduced in the EFT description, especially near the \(H_2\) pole. At the same time, this viable space is localized and will be significantly probed by future experiments such as DARWIN.

\section{Conclusions}
\label{sec:conclusions}

In this work we have revisited the Higgs-portal scenario for vector DM, comparing its EFT description with a renormalizable UV-complete realization. Our goal was to assess to what extent the phenomenological conclusions inferred from the EFT remain valid once the model is embedded in a consistent gauge theory.

In the EFT approach, the DM particle is described as a massive Proca vector field \(V_\mu\) coupled to the SM through the operator \((H^\dagger H)V_\mu V^\mu\). This framework is economical and convenient for phenomenological studies, since it is effectively characterized by only two parameters, the DM mass \(m_V\) and the Higgs-portal coupling \(\lambda_{HV}\). We have shown, however, that this description has a limited range of validity because of perturbative unitarity at high energies, and is already under very strong experimental pressure. Combining the relic-density requirement with limits from DD, ID, and invisible Higgs decays, we find that the EFT parameter space is essentially excluded, except for a very narrow region close to the Higgs resonance, \(m_V \simeq m_h/2\), which requires a fine tune of the DM mass at the permille level that is theoretically difficult to explain. Even in this case, the viable solutions rely on resonantly enhanced annihilation and are expected to be decisively tested by future DD experiments such as DARWIN.

We then considered a UV-complete realization based on a gauged \(U(1)_X\) symmetry. In this setup, the dark sector contains the vector boson \(V\) and a complex scalar \(S\), whose vacuum expectation value spontaneously breaks \(U(1)_X\) and generates the DM mass. A \(\mathbb{Z}_2\) symmetry forbids kinetic mixing with hypercharge and ensures the stability of the dark vector. After symmetry breaking, the scalar sector contains two CP-even states, the SM-like Higgs \(H_1\) and a second scalar \(H_2\), mixed by an angle \(\theta\). The presence of this second mediator qualitatively modifies the phenomenology with respect to the EFT description.

Our analysis shows that the most important new feature of the UV-complete model is the presence of an additional resonance at \(m_V \simeq m_{H_2}/2\). Near this pole, annihilation in the early Universe can be significantly enhanced, allowing the observed relic density to be reproduced with much smaller values of the portal coupling \(\lambda_{HS}\) than would be required away from resonance. For sufficiently small scalar mixing, this mechanism opens viable thermal-relic parameter space that is absent in the EFT case. In particular, for representative benchmarks such as \(\sin\theta=0.05\), i.e.~in the small mixing case, the coupling required by the relic-density constraint near the \(H_2\) resonance can lie well below future DARWIN DD bounds and, depending on the benchmark, up to about two orders of magnitude below the present LZ limits. The surviving solutions are nevertheless localized around the resonance and therefore remain moderately tuned, in the sense that \(m_V\) must lie sufficiently close to \(m_{H_2}/2\), typically at the \(\mathcal{O}(10\%)\) level. This is comparatively less severe than the per-mille-level mass tuning required in the EFT case, and may be easier to accommodate in a UV-complete framework.

At the same time, the UV-complete model is subject to additional constraints. For \(m_V<m_{H_1}/2\), invisible Higgs decays constrain the low-mass region. Moreover, large scalar mixing angles are generally in tension with Higgs signal-strength measurements and with direct searches for additional scalar states. As a result, the phenomenologically most robust region of parameter space is the one with relatively small \(\sin\theta\), where the \(H_2\) resonance is narrow enough to suppress DD while still maintaining efficient annihilation at freeze-out.

Overall, our results highlight a central lesson: for vector Higgs-portal dark matter, the EFT and UV-complete descriptions can lead to qualitatively different conclusions. While the EFT suggests that the model is almost entirely ruled out, the renormalizable \(U(1)_X\) completion retains viable thermal-relic regions thanks to the additional scalar mediator and the associated resonant dynamics. This shows that a consistent UV completion is not only theoretically well motivated, but also essential for a reliable interpretation of present and future searches for Higgs-portal vector DM.

Future progress will come from several complementary directions. Improved DD sensitivity, especially from DARWIN-like experiments, will probe a large fraction of the surviving parameter space. At the same time, more precise measurements of Higgs properties and dedicated searches for additional scalar resonances at colliders will further constrain the mixing structure of the model. Together, these probes will determine whether the resonant UV-complete Higgs-portal scenario can remain a viable explanation of DM.

\begin{acknowledgments}
H.S. acknowledges support from the Collaborative Research Center SFB1258 and from the Deutsche Forschungsgemeinschaft (DFG, German Research Foundation) under Germany's Excellence Strategy - EXC-2094 - 390783311. M.D.M. acknowledges support from the research grant {\sl TAsP (Theoretical Astroparticle Physics)} funded by Istituto Nazionale di Fisica Nucleare (INFN). M.D.M. acknowledges support  from the Italian Ministry of University and Research (MUR), PRIN 2022 ``EXSKALIBUR – Euclid-Cross-SKA: Likelihood Inference Building for Universe’s Research'', Grant No. 20222BBYB9, CUP I53D23000610 0006, and from the European Union -- Next Generation EU.
\end{acknowledgments}

\bibliographystyle{apsrev4-2}
\bibliography{paper}

\end{document}